\documentclass[onecolumn,trackchanges]{aastex62}

\usepackage{graphicx}
\usepackage{subfigure}
\submitjournal{ApJ}
\newcommand{\rhessi}{\textit{RHESSI}}
\newcommand{\foxsi}{\textit{FOXSI-2}}
\newcommand{\fox}{\textit{FOXSI}}
\newcommand{\aia}{\textit{SDO}/AIA}
\newcommand{\xrt}{\textit{Hinode}/XRT}
\newcommand{\goes}{\textit{GOES}}
\newcommand{\nustar}{\textit{NuSTAR}}

\newcommand{\magixs}{\textit{MaGIXS}}

\shortauthors{Athiray et al.}
\shorttitle{{\foxsi} Microflares DEM}

\begin{document}



\title{{\foxsi} Solar Microflares I : Multi-instrument Differential  Emission Measure Analysis and Thermal Energies}

\correspondingauthor{P.S. Athiray}
\email{athiray.panchap@nasa.gov}

\author[0000-0002-4454-147X]{P.S. Athiray}
\affiliation{University of Minnesota, Twin Cities, Minneapolis}
\affiliation{NASA Postdoctoral Program, NASA Marshall Space Flight Center, ST13, Huntsville, AL}

\author[0000-0002-7407-6740]{Juliana Vievering}
\affiliation{University of Minnesota, Twin Cities, Minneapolis}

\author[0000-0001-7092-2703]{Lindsay Glesener}
\affiliation{University of Minnesota, Twin Cities, Minneapolis}

\author{Shin-nosuke Ishikawa}
\affiliation{Nagoya University, Japan}

\author{Noriyuki Narukage}
\affiliation{National Astronomical Observatory of Japan}

\author[0000-0002-8203-4794]{Juan Camilo Buitrago-Casas}
\affiliation{Space Sciences Laboratory, University of California at Berkeley}

\author[0000-0002-0945-8996]{Sophie Musset}
\affiliation{University of Minnesota, Twin Cities,  Minneapolis}

\author[0000-0003-0656-2437]{Andrew Inglis}
\affiliation{The Catholic University of America}

\author[0000-0001-6127-795X]{Steven Christe}
\affiliation{NASA Goddard Space Flight Center}

\author{S\"am Krucker}
\affiliation{University of Applied Sciences and Arts Northwestern Switzerland}
\affiliation{Space Sciences Laboratory, University of California at Berkeley}

\author{Daniel Ryan}
\affiliation{NASA Goddard Space Flight Center}

\begin{abstract}
In this paper we present the differential emission measures (DEMs) of two sub-A class microflares observed in hard X-rays (HXRs) by the {\foxsi} sounding rocket experiment, on 2014 December 11. The second {\fox} ({\it Focusing Optics X-ray Solar Imager}) flight was coordinated with instruments  {\xrt} and {\aia}, which provided observations in soft X-rays (SXR) and Extreme Ultraviolet (EUV). This unique dataset offers an unprecedented temperature coverage useful for characterizing the plasma temperature distribution of microflares. By combining data from {\foxsi}, XRT, and AIA, we determined a well-constrained DEM for the microflares. The resulting DEMs peak around 3MK and extend beyond 10MK.  The emission measures determined from {\foxsi} were lower than 10$^{26} cm^{-5}$ for temperatures higher than 5MK; faint emission in this range is best measured in HXRs. The coordinated {\foxsi} observations produce one of the few definitive measurements of the distribution and the amount of plasma above 5MK in microflares. We utilize the multi-thermal DEMs to calculate the amount of thermal energy released during both the microflares as ${\sim}$ 5.0 ${\times}$ 10$^{28}$ ergs for Microflare 1 and ${\sim}$ 1.6 ${\times}$ 10$^{28}$ ergs for Microflare 2. We also show the multi-thermal DEMs provide a more comprehensive thermal energy estimates than isothermal approximation, which systematically underestimates the amount of thermal energy released.
\end{abstract}
\keywords{{\fox}, Solar microflares, Differential Emission Measure}
\section{Introduction}
\label{sec:intro}

Solar flares exhibit a wide range of temperatures that emit over a wide energy range. Joint observations in the extreme ultraviolet (EUV), soft X-rays (SXRs) and hard X-rays (HXRs) have been useful to determine the underlying temperature distribution of the hot coronal plasma \citep[e.g.][]{Ishikawa2017,Wright2017,Ishikawa2019}. Current space instrumentation in the EUV and SXRs have limited sensitivity to plasma above 5MK \citep{Winebarger2012}. HXRs have the ability to better  constrain the temperature distribution at these temperatures. As of today, much of our knowledge on the high energy aspects of solar flares is derived from the {\it Reuven Ramaty High Energy Solar Spectroscopic Imager} ({\rhessi}) \citep{Lin2002}. The {\rhessi} instrument was sensitive to high temperature plasma and observed large scale flares and microflares.  However, {\rhessi} had limited instrument sensitivity for small active regions due to indirect imaging and high detector background.

The distribution of flare frequency suggests that smaller energy releases occur more often than large flares. Multiple  past studies find that the  flare frequency distribution follows a negative power law relation with an index ${\approx}$ 2 \citep[e.g.][]{Hudson1991,Hannah2008,Hannah2011,Aschwanden2012}.  Decades of {\rhessi} observations including more than 20,000 flares have been useful for studying the frequency of microflares and associated energetics, as reported in \citet{Christe2008} and \citet{Hannah2008}.   However, it is not clear how to keep the corona consistently at high temperatures with small flare events. Therefore, it is important to investigate ``small-scale" energy releases (fainter than {\rhessi} observations) at high temperatures to understand the plasma temperature distribution, amount of energy released, and how they compare with larger flares. Higher sensitivity and dynamic range can be achieved by using direct focusing X-ray optics, which provide a monotonically falling point spread function (PSF) and a high signal-to-noise ratio due to  smaller detector area. The {\it Nuclear Spectroscopic Telescope Array} (\nustar) is the  first satellite mission to use direct focusing optics in HXRs (3 -  79 keV) \citep{Harrison2013}, designed specifically for astrophysical observations. The {\nustar} solar campaigns have been useful in studying small-scale energy releases including active regions, microflares, and quiet Sun flares \citep[e.g.][]{Hannah2016, Glesener2017, Wright2017, Kuhar2018}. However, the observations are restricted due to relatively low detector live time for even small-scale events as the instrument design is not optimized for solar observations.

The {\it Focusing Optics X-ray Solar Imager} (\fox) \citep{Krucker2014} is a sounding rocket experiment funded by NASA's Low Cost Access to Space program. {\fox} is the first HXR imaging spectroscopy experiment dedicated for solar observations to use direct focusing optics in the energy range 4 to 20 keV. {\fox} demonstrates the unique power of direct focusing optics to achieve sensitivity $\gtrapprox$ 10 times greater than that of {\rhessi}. So far, {\fox} has had three successful rocket flights, in the years 2012, 2014, and 2018. The second flight ({\foxsi}) carried seven Wolter-I type optic modules with a 2m focal length paired with seven semiconductor detectors made of Si and CdTe \citep{Glesener2016,Christe2016}. {\foxsi} was sensitive to plasma emission above 5MK, critical for the measurement of high temperature emission from active regions and microflares.

In this paper (Paper I), we present a comprehensive differential emission measure (DEM) analysis of two sub-A class microflares jointly observed by {\foxsi}, {\xrt} and {\aia}. We determine the plasma temperature distribution of microflares and present
comprehensive estimates of thermal energy released from the microflares using  multi-thermal DEMs. The high sensitivity {\foxsi} data also allow us to perform a detailed imaging and spectroscopic analysis of the microflares in HXRs, which are presented in Vievering et al. (2019 under preparation) (here onwards Paper II). Paper II also describes thermal energy estimates using an isothermal approximation and explores the energy of non-thermal  electrons in the microflares. For this paper, the coordinated {\foxsi} observations are described in Section \ref{label:coordinatedobs}. In Section \ref{label:tempresp}, we explain the construction of the temperature response function for {\foxsi}. Combined DEM analyses and thermal energy estimates are discussed in Section \ref{label:DEManalysis} and  \ref{sec:Ten_estimates}. Finally, a summary of the investigation is presented in Section \ref{label:summary}.

 \section{{\foxsi} and its coordinated observations}
 \label{label:coordinatedobs}
 {\foxsi} was successfully launched on 11 December 2014 from the White Sands Missile Range. It observed the Sun for 6 minutes and 41 seconds starting from 19:12:42 UT. The observations included five targets that covered many interesting features including two microflares, three  quiescent active regions (AR) and some portions of the quiet Sun. A list of all targets observed during the {\foxsi} flight is given in Paper II. Here, we focus on targets for which microflares are observed as listed in Table~\ref{tab:foxsiobstimes}. This flight was coordinated with several other instruments including the X-ray Telescope (XRT) onboard {\it Hinode} \citep{Golub2007}. We also exploit data from the Atmospheric Imaging Assembly (AIA) onboard the Solar Dynamics Observatory ({\it SDO}) \citep{Lemen2012}, which observes the full Sun all the time.

\subsection{{\foxsi} instrument and data description}
\label{label:foxsiinstrument}

{\foxsi} carried seven direct-focusing optics modules, each paired with a dedicated photon-counting double-sided semiconductor strip detector.
Five out of the seven optics modules had 7 nested Wolter-I mirrors, while the other two optics modules had 10 nested Wolter-I mirrors in each. The optics modules were produced using an electroformed nickel replication process at the Marshall Space Flight Center (MSFC) \citep{Ramsey2005}. The double-sided strip detectors were composed of Si and CdTe with pitches\footnote{Pitch - Distance between the centers of two adjacent strips} of 75 ${\mu}$m and 60${\mu}$m, respectively. With the 2m focal length of the {\fox} optics, the Si detectors had an angular resolution of 7.7 arcseconds, while the CdTe detectors had an angular resolution of 6.2 arcseconds. The active volume of the detector carried 128 orthogonal strips on each side of the detector, which give a two-dimensional position for each photon interaction. The arrangement of 128 ${\times}$ 128 strips yield 16,384 strip crossings, which is the smallest individual element in an image (analogous to `{\it 1 pixel}' in pixelated detectors).  The experiment was designed to observe in the energy range 4 - 20 keV. Each optic/detector pair collected an independent measurement. The processed level 2 data contained information on the photon energy and the position of each photon interaction on the detector. X-ray images are produced in solar cartesian coordinates for different energy and time intervals by translating from photon-hit detector coordinates using the geometry and orientation of each detector. For a given energy range, each point in the image represents the intensity measured in counts s$^{-1}$.

\begin{figure}[h]
\hspace{-4.5em}\includegraphics[width=0.45\linewidth]{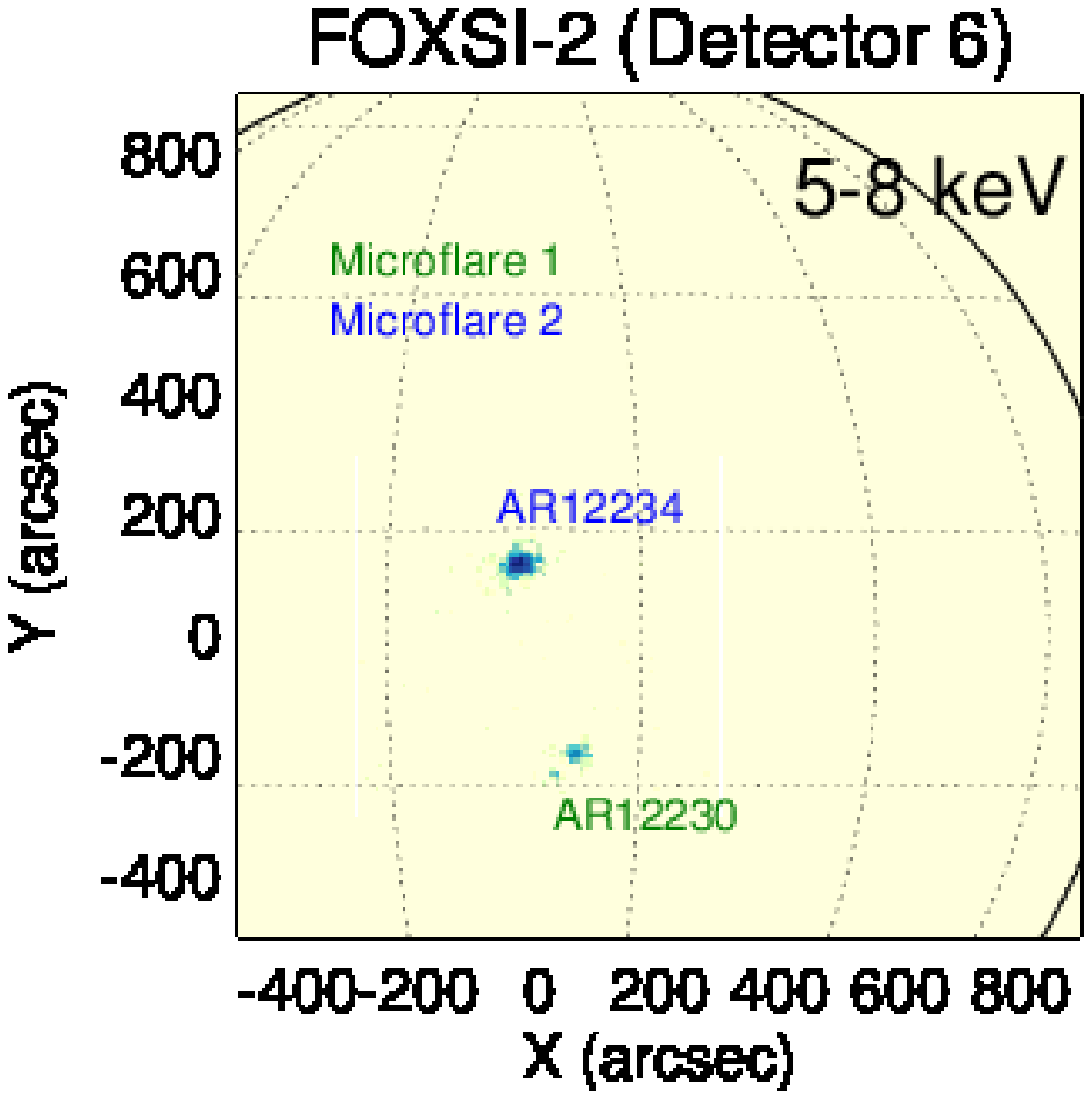}
\hspace{-7.6em}
\includegraphics[width=0.46\linewidth]{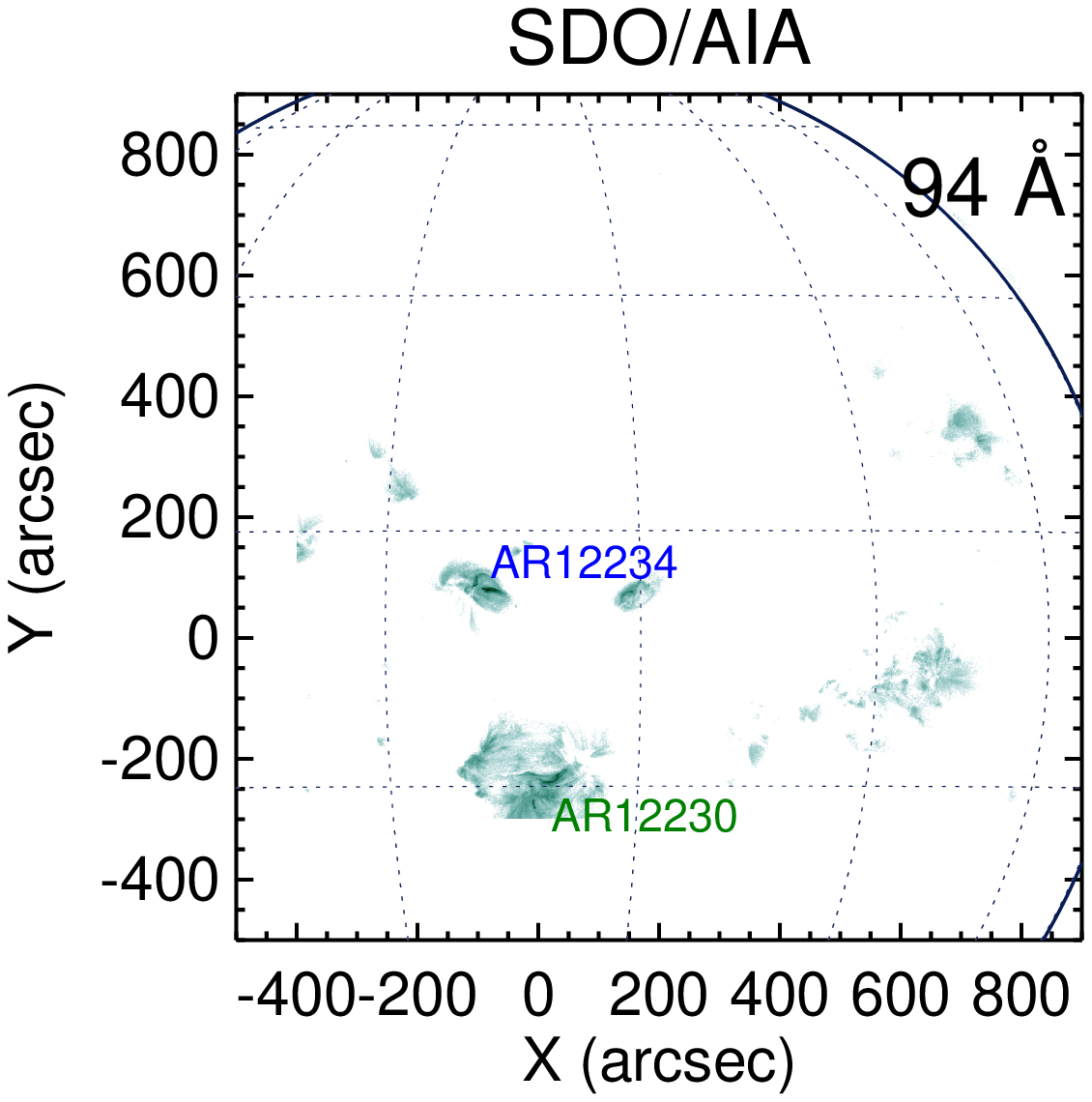}
\hspace{-7.6em}
\includegraphics[width=0.46\linewidth]{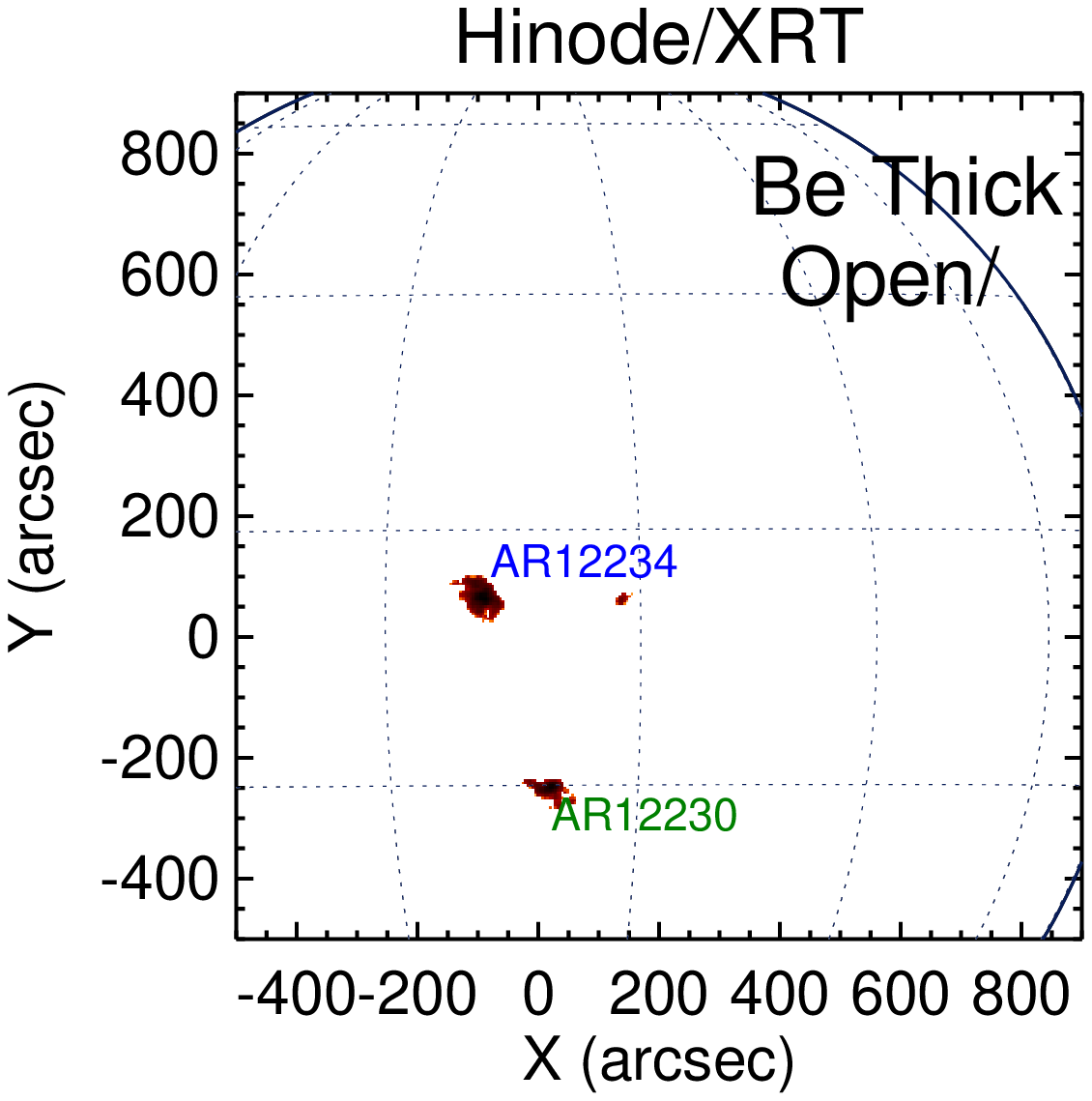}\hfill
\caption{(Left){\foxsi} X-ray image of AR12230 and AR12234 in 5 - 8 keV from a Si detector (D6). Microflares were observed from these  ARs at different times during the {\foxsi} flight.
This image corresponds to a time interval from 19:18:51 to  19:19:23 UT (Target J). Simultaneous observation of both the ARs (one flaring and one not) within the same FOV demonstrates the capability of {\foxsi} to image sources of different intensities.(Middle and Right) Corresponding {\aia} 94 {{\AA}} and {\xrt} images of the same ARs.}

  \label{fig:Microflaresimage_fov}
\end{figure}

\begin{figure}[h]
\hspace{-5em}
\includegraphics[width=0.65\linewidth]{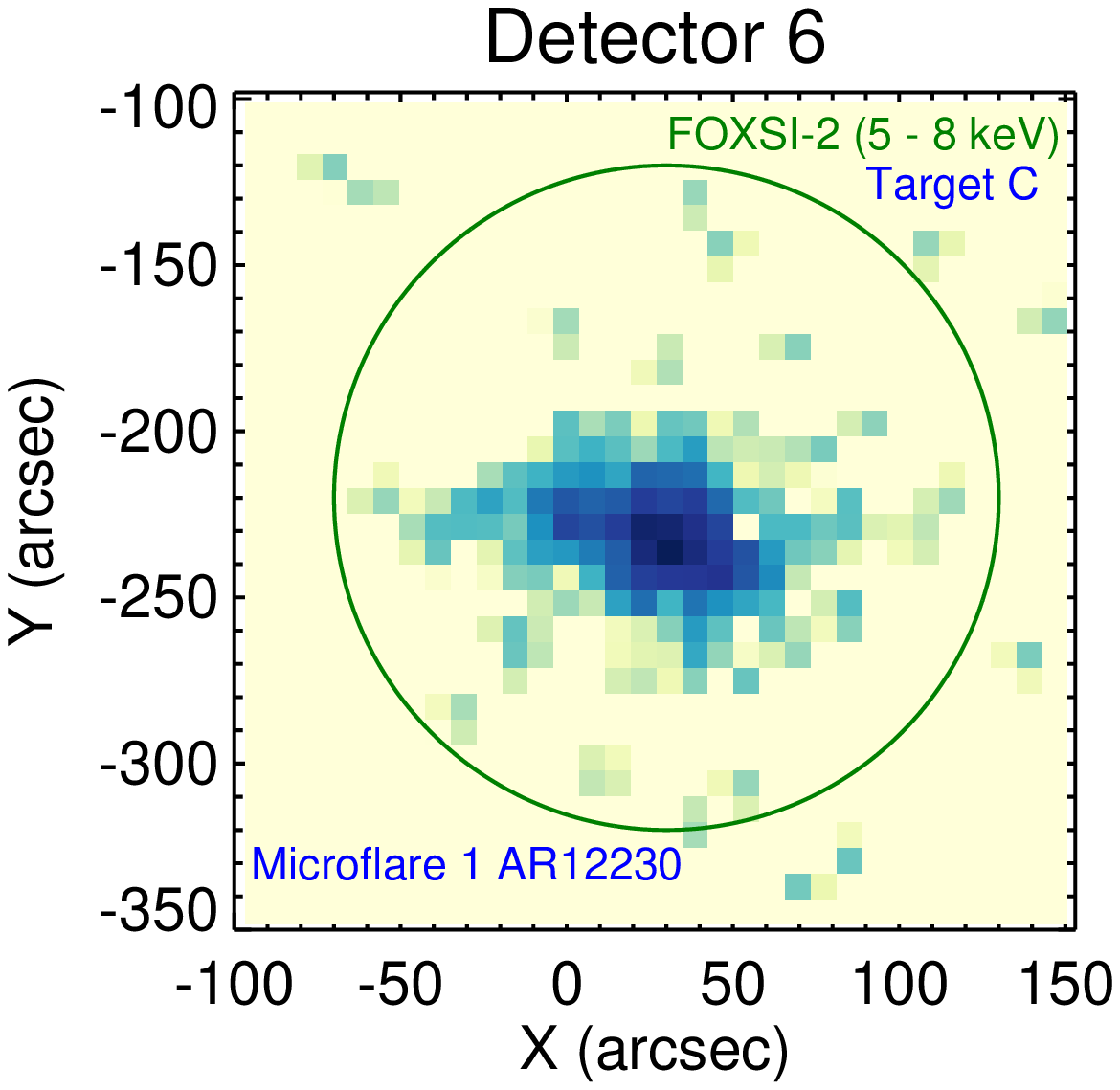}
\hspace{-6.2em}
\includegraphics[width=0.59\linewidth]{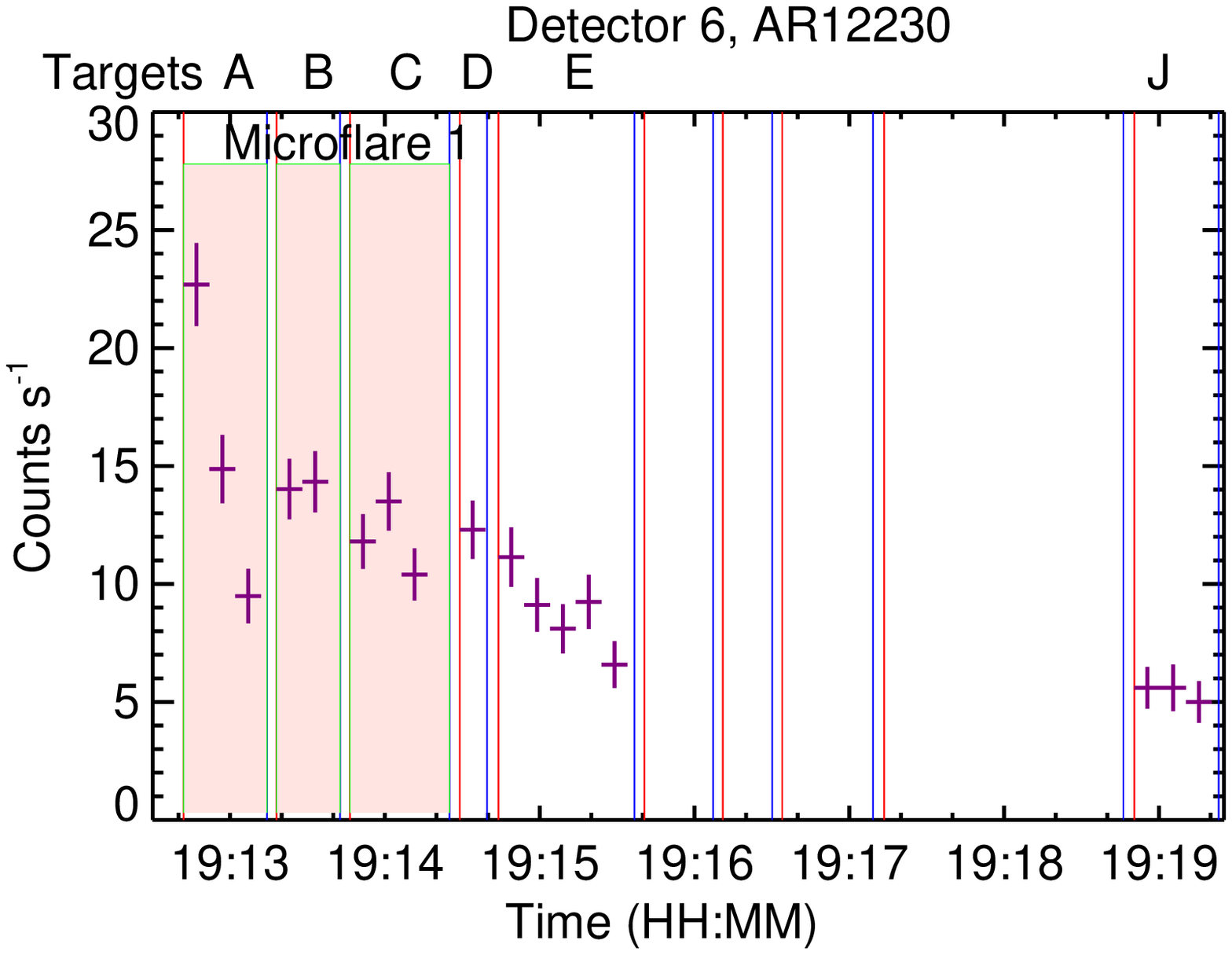}
\vspace{-1em}

\hspace{-5em}
\includegraphics[width=0.65\linewidth]{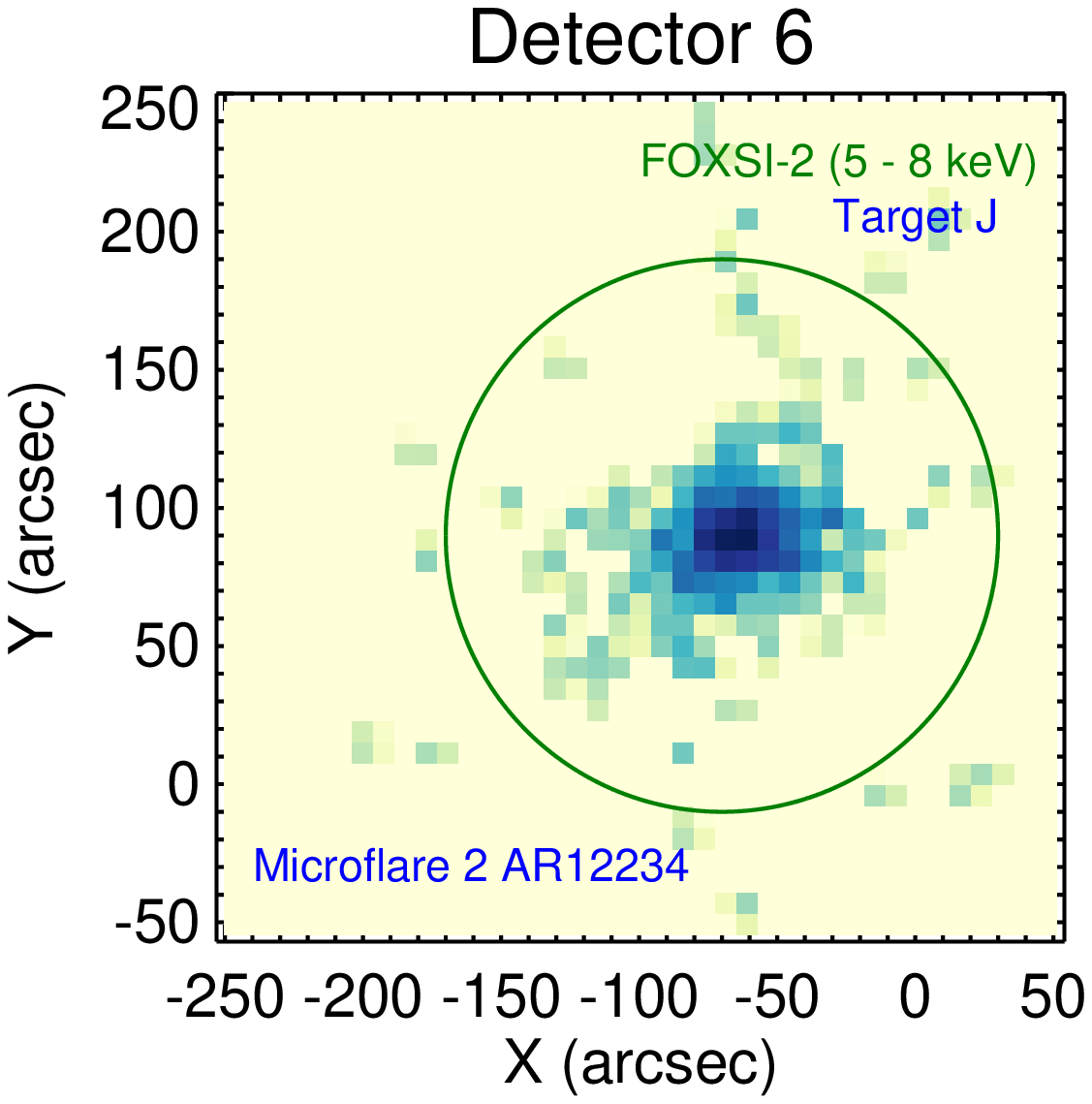}
\hspace{-6.2em}
\includegraphics[width=0.59\linewidth]{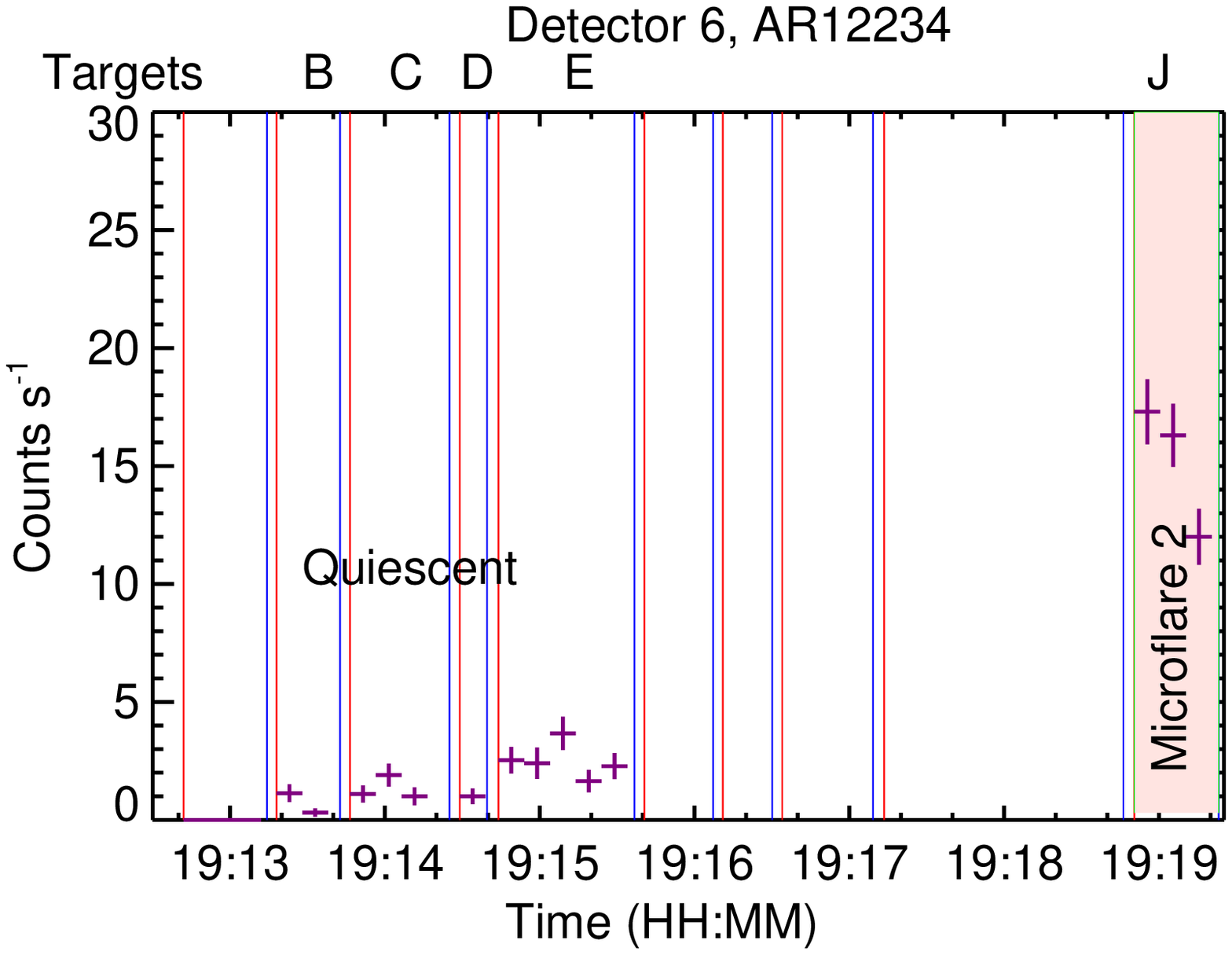}

 \caption{(Top) {\foxsi} HXR image (Left) and temporal profile (Right) of AR12230 in the 5 - 8 keV energy range from a Si detector (D6). This image corresponds to Target C  (19:13:47 to 19:14:25 UT). (Bottom) HXR image (Left) and temporal profile (Right) of AR12234 in the 5 - 8 keV energy range from a Si detector (D6). This image corresponds to Target J (19:18:51 to 19:19:23 UT). The circles (radius = 100 arcseconds) indicate the regions chosen for flux determination for DEM analysis. The vertical bars in the light curve denote target changes, which were fine adjustments to center the desired target in the FOV. Observations in the shaded time intervals (rose color) corresponding to the Targets A, B, C, and J are discussed in this paper.}
  \label{fig:Microflareimage}
\end{figure}

\subsection{{\foxsi} microflare observations}
\label{label:foxsisummary}
Table \ref{tab:foxsiobstimes} summarizes the targets observed by {\foxsi} in which Microflare 1 and Microflare 2 were observed. The two microflares observed by {\foxsi} occurred in different ARs. Microflare 1 was observed from NOAA AR12230; Microflare 2 was observed from NOAA AR12234. The separation between these ARs (${\sim}$ 3.0 arcminutes) is such that {\foxsi} can distinctly observe them within its  FOV of 16.5 arcminutes as shown in Figure \ref{fig:Microflaresimage_fov} (Left).  This image is obtained during Target J (19:18:51 to 19:19:23 UT) using a Si  detector (D6). During the Targets B through E, both the ARs were observed inside {\foxsi}'s FOV. Just before the end of flight {\foxsi} observed both the ARs during Target J for 32 seconds from 19:18:51 UT. During this time, an attenuator was inserted in the optical path in front of six out of the seven detectors  (all but D6). Simultaneous observation of both the ARs within the same FOV demonstrates the capability of {\foxsi} to image sources with different HXR intensities.  HXR images of the microflares from {\foxsi} at different targets are shown in Paper II.  Context data for the ARs in the EUV and SXRs are obtained from {\aia} and {\xrt} as shown in  Figure~\ref{fig:Microflaresimage_fov} (Middle and Right), and a description is given in Section~\ref{label:aiaxrtsummary}.

\setcounter{table}{0}
\begin{table}[h!]
\renewcommand{\thetable}{\arabic{table}}
\centering
\caption{Targets observed by {\foxsi} that include Microflare 1 and Microflare 2} \label{tab:foxsiobstimes}
\begin{tabular}{ccccccc}
\tablewidth{0pt}
\hline
\hline
Flare& Target&Start & End & Duration & Target center & {\xrt} filters \\
&&(UT) & (UT) & (sec) & (arcsec)& (Filter 1/Filter 2) \\
\hline
1&A&19:12:42 & 19:13:14 &32 & 359,-431& C poly/Ti poly \\
	&	 && & & &C poly/Open\\
         \hline
1&B&19:13:18 & 19:13:42 &24 & -1 ,-431& Be thin/Open \\
&&&&&&Be med/Open\\
\hline
1&C&19:13:47 & 19:14:25 &38 & -1, -251& Be med/Open \\
&&&&&&Al med/Open\\
\hline
1&D&19:14:29 & 19:14:39 &10 & -1,-281& Al med/Open\\
\hline
1&E&19:14:44 & 19:15:36 &52 &  -390, -281& Al poly/Open  \\
\hline
2&J&19:18:51 & 19:19:23 &32 & 0, -251& Open/Be thick \\
\hline
\end{tabular}
\end{table}

We first grouped {\foxsi} data into equal width energy bins, with 1 keV bin size. We then selected only bins with ${\geq}$ 15 to 20 photons, which resulted in the energy range 4 to 8 keV. For energies below 5 keV, the knowledge of detector's spectral response is less well-characterized due to the uncertainty in the quantum efficiency curve near the low energy threshold, programmed in the ASIC to be at 4 keV \citep{Lindsay2012}. This affects the precise estimation of the incident flux at low energy X-rays and limits the use of data below 5 keV. Therefore, we limit our analysis to data from 5 to 8 keV. Figure \ref{fig:Microflareimage} (Left) shows HXR images of Microflare 1 and Microflare 2 in the 5 - 8 keV energy range at different time intervals. The HXR image was obtained from one of the Si detectors (D6), paired with a 10 shell optic that had the highest effective area.   The methodology to determine flare centroid in {\foxsi} images by co-aligning with {\rhessi} data is described in Paper  II. The circles (radius = 100 arcseconds) indicate the regions chosen to obtain flux for the DEM analysis.  The corresponding time evolution is shown in Figure \ref{fig:Microflareimage} (Right). Different targets are denoted in the light curve as vertical bars with  colors red and blue indicating the beginning and end of each target,  respectively.  The light curve of  Microflare 1 shows a decrease in HXR counts with time. For Microflare 2, an increase in the HXR counts is observed during Target J.  The HXR count flux for AR12234 stays  below 10 counts/s during non-flaring times, labelled in Figure \ref{fig:Microflareimage} (Bottom Right). The complexity of these flaring ARs is studied using HXR imaging spectroscopy in Paper II. This reveals  spatial variation of low- and high-energy HXR emission from within the bright loop, showing evidence of a multi-thermal plasma.

\subsection{{\aia} and {\xrt} event overview}
\label{label:aiaxrtsummary}

The EUV images of microflares were obtained from {\aia} with 1.2 arcsecond angular resolution and 12 seconds cadence \citep{Lemen2012}. We have considered five EUV channels  (94 {{{{\AA}}}}, 131 {{{{\AA}}}}, 171 {{{{\AA}}}},  193 {{{{\AA}}}}, and 211 {{{{\AA}}}}) which are sensitive to temperatures above log T = 5.6 for our investigation. We have excluded the AIA 335 {{{{\AA}}}} channel due to a long-term drop in sensitivity resulting from contamination accumulation \citep{Boerner2014}. {\xrt} performed coordinated observations as requested by the {\foxsi} team using the observation plan HOP 221 for DEM investigation using multiple filters. {XRT} images were taken with multiple filters with a 4 arcsecond ${\times}$4 arcsecond resolution (ie., 4${\times}$4 CCD pixel binning). During the {\foxsi} flight, the field of view of {\xrt} covered active regions near the center of the solar disk, including both the microflares. A list of XRT filters used in our DEM analysis is given in Table \ref{tab:foxsiobstimes}. Both  AIA and XRT data are processed using standard SolarSoft procedures (aia\_dataprep.pro and xrt\_prep.pro).

\begin{figure}[h]
\hspace{-1.3em}
\includegraphics[width=0.28\linewidth,trim ={2cm 1cm 3cm 0cm},clip]{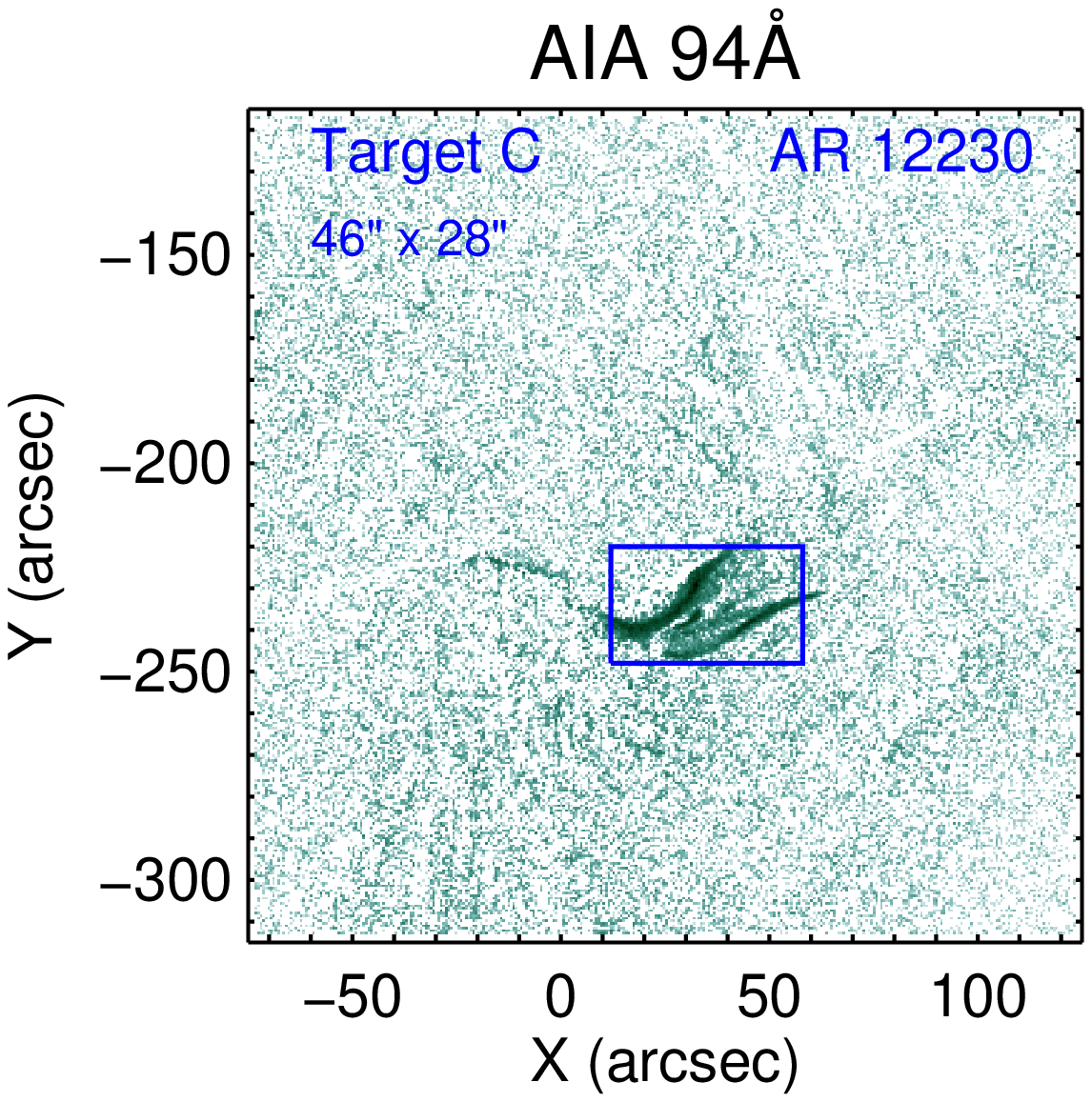}
\hspace{-1.5em}
\includegraphics[width=0.26\linewidth,trim ={3cm 1cm 3cm 0cm},clip]{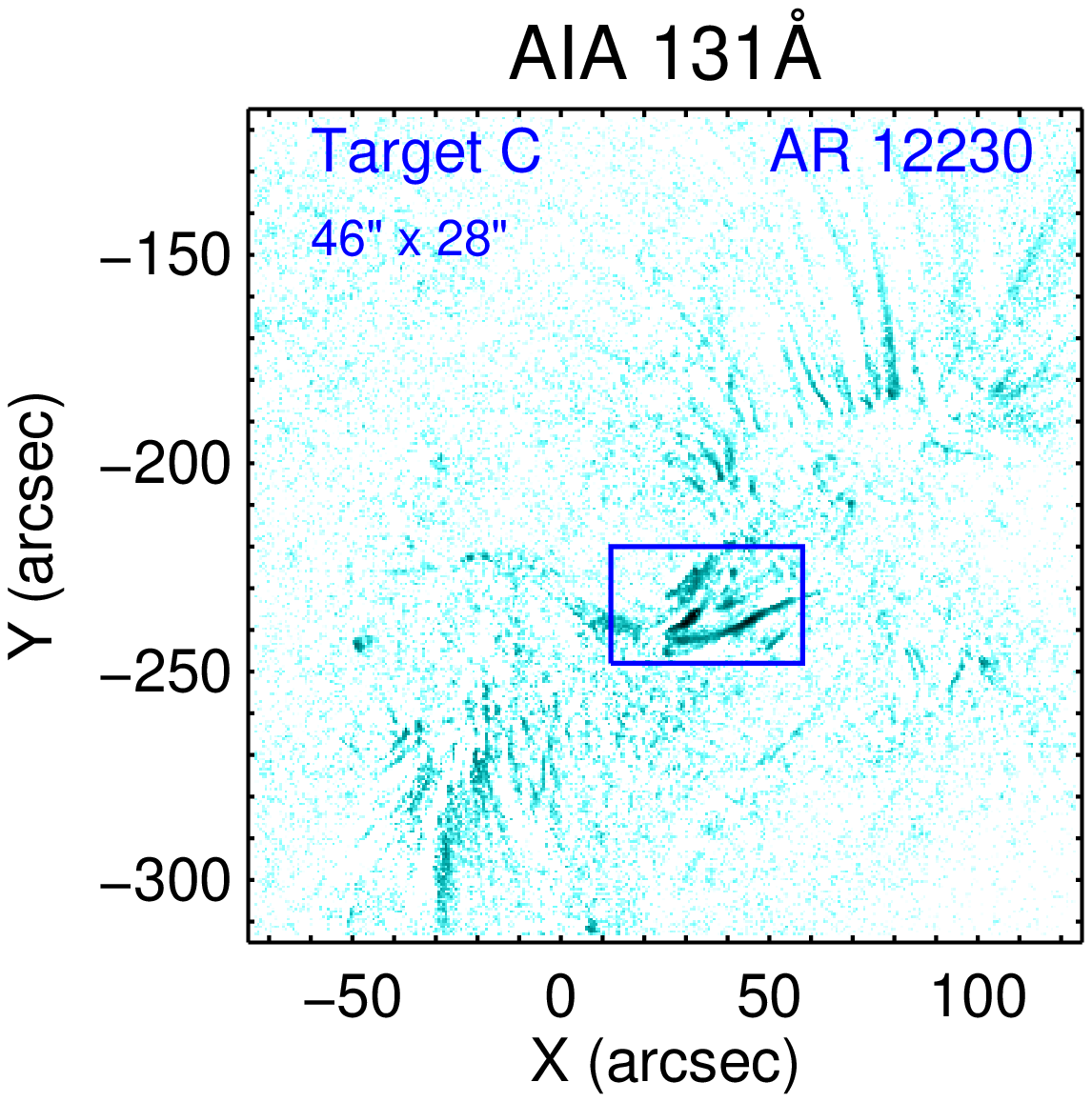}
\hspace{-1.5em}
\includegraphics[width=0.26\linewidth,trim ={3cm 1cm 3cm 0cm},clip]{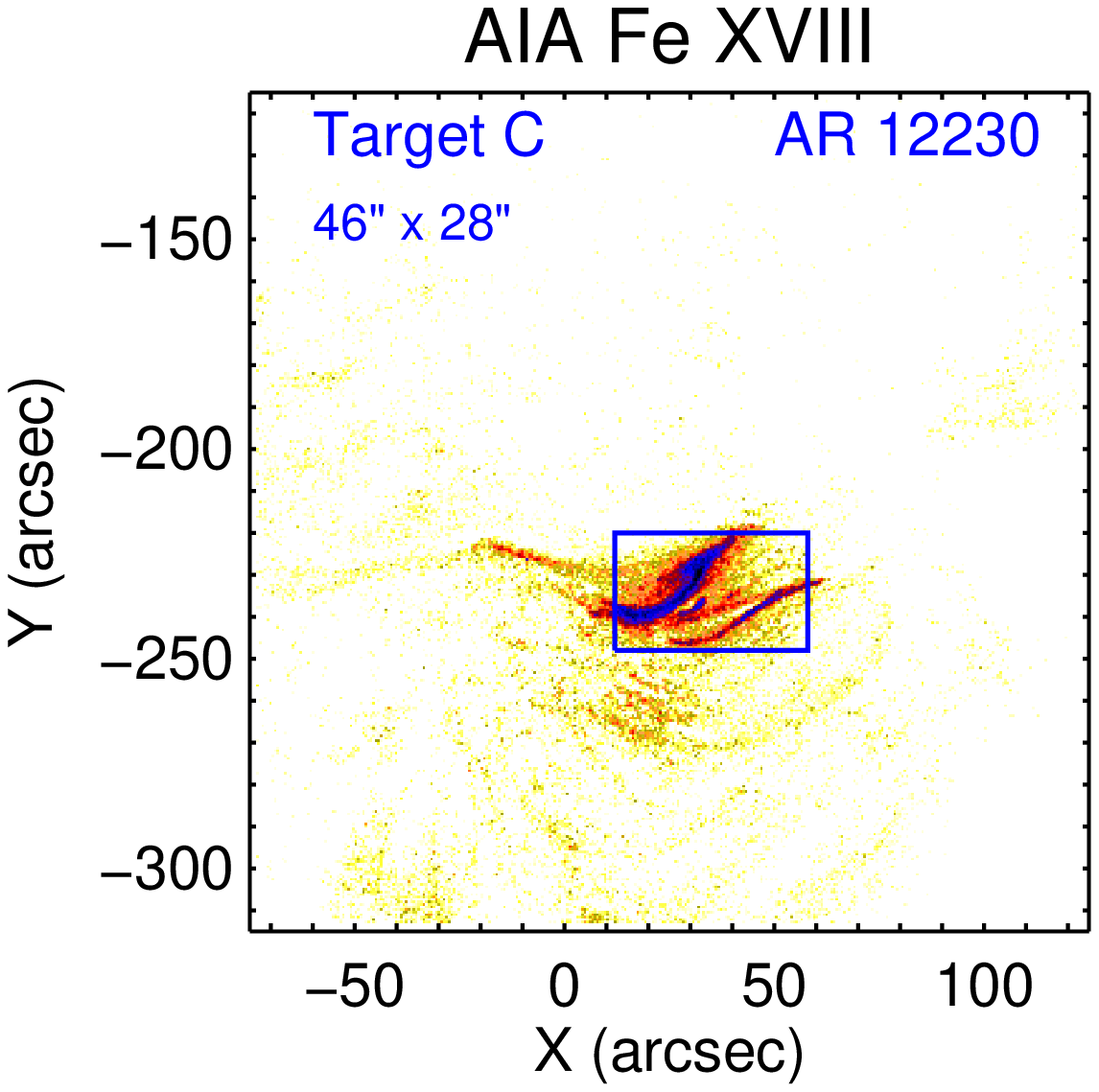}
\hspace{-1.5em}
\includegraphics[width=0.26\linewidth,trim ={3cm 1cm 3cm 0cm},clip]{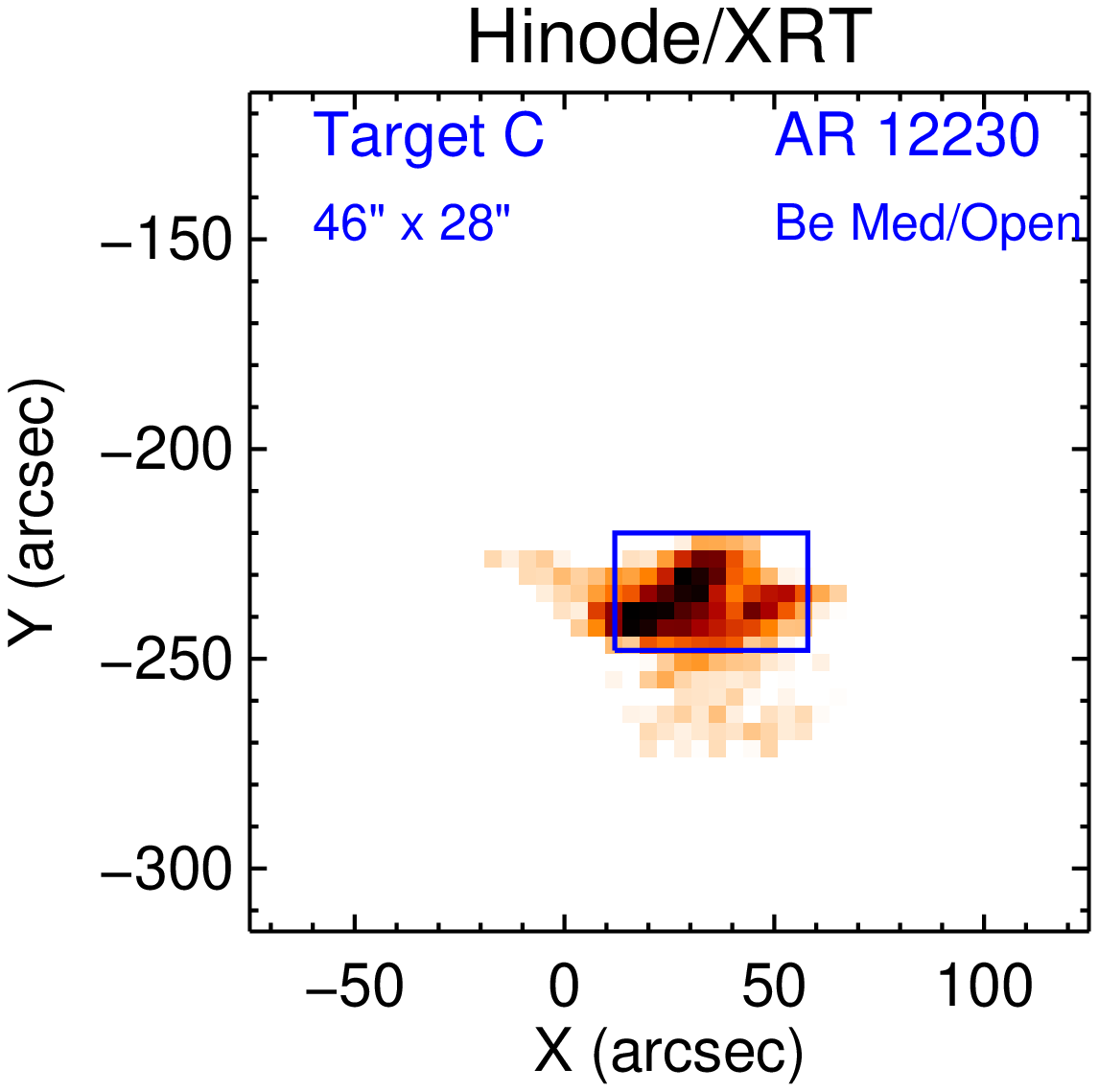}
\vspace{0.40mm}
\hspace{-1.3em}
\includegraphics[width=0.28\linewidth,trim ={2cm 0cm 3cm 1.65cm},clip]{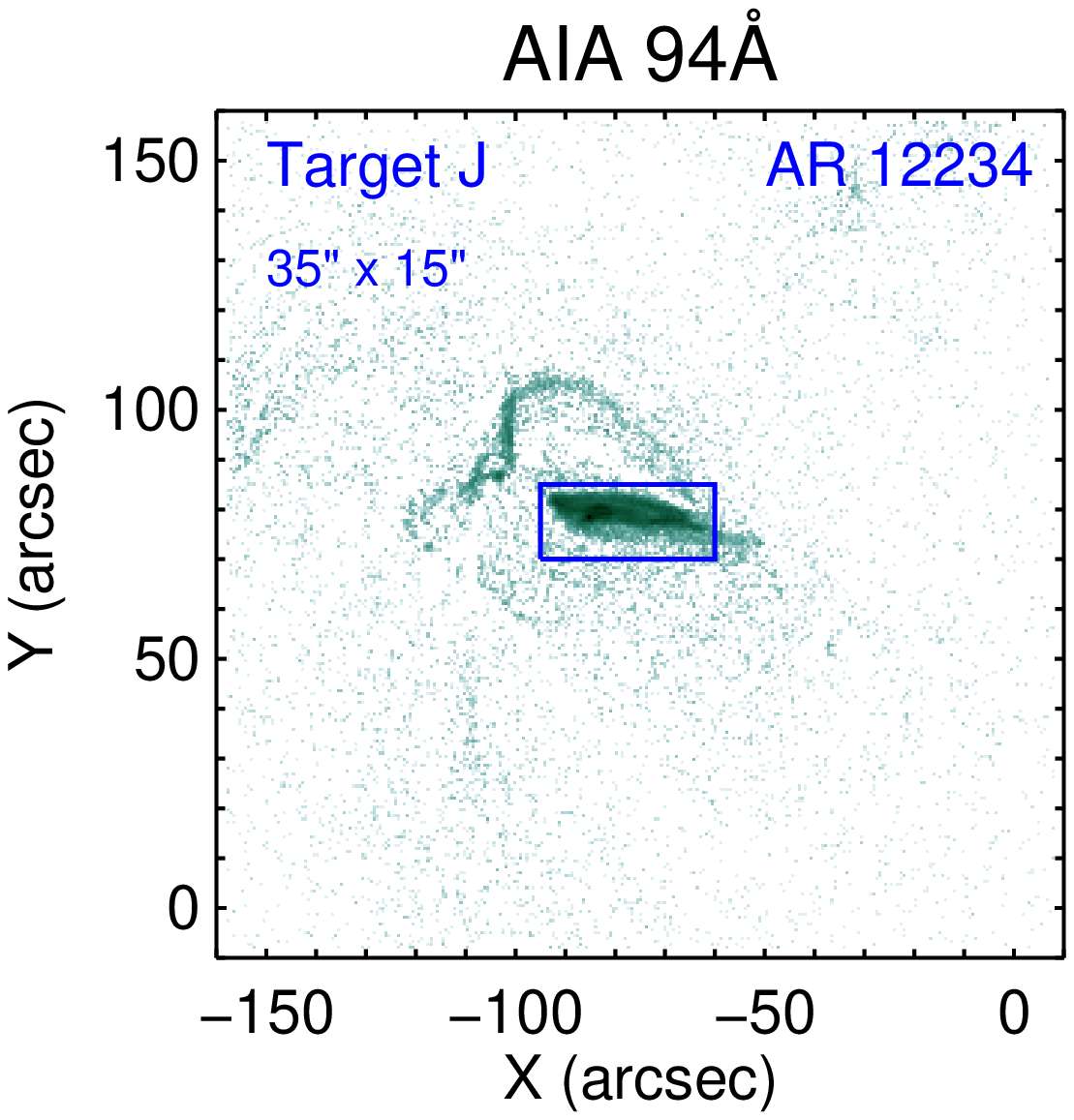}
\hspace{-1.5em}
\includegraphics[width=0.26\linewidth,trim ={3cm 0cm 3cm 1.65cm},clip]{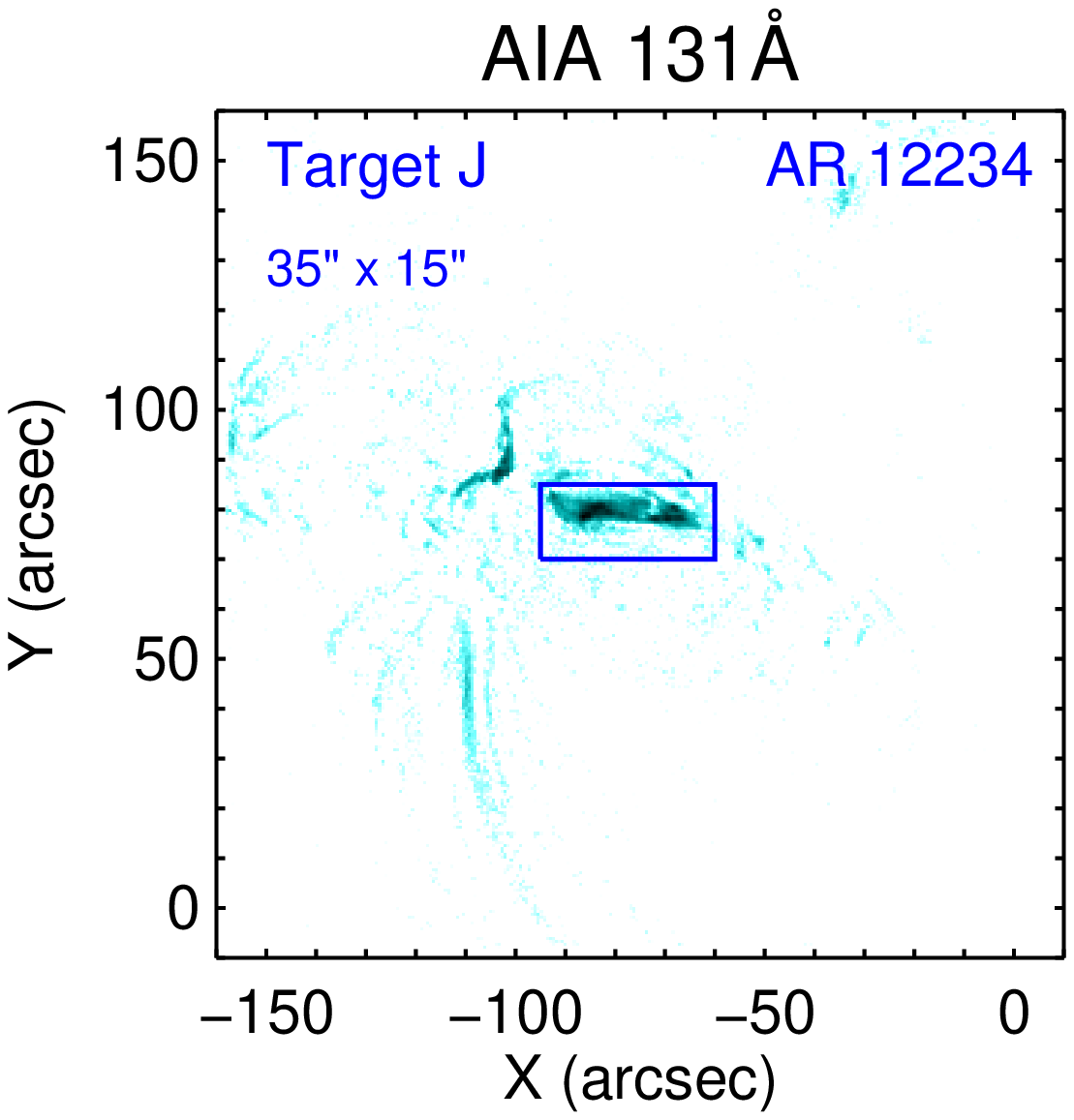}
\hspace{-1.5em}
\includegraphics[width=0.26\linewidth,trim ={3cm 0cm 3cm 1.65cm},clip]{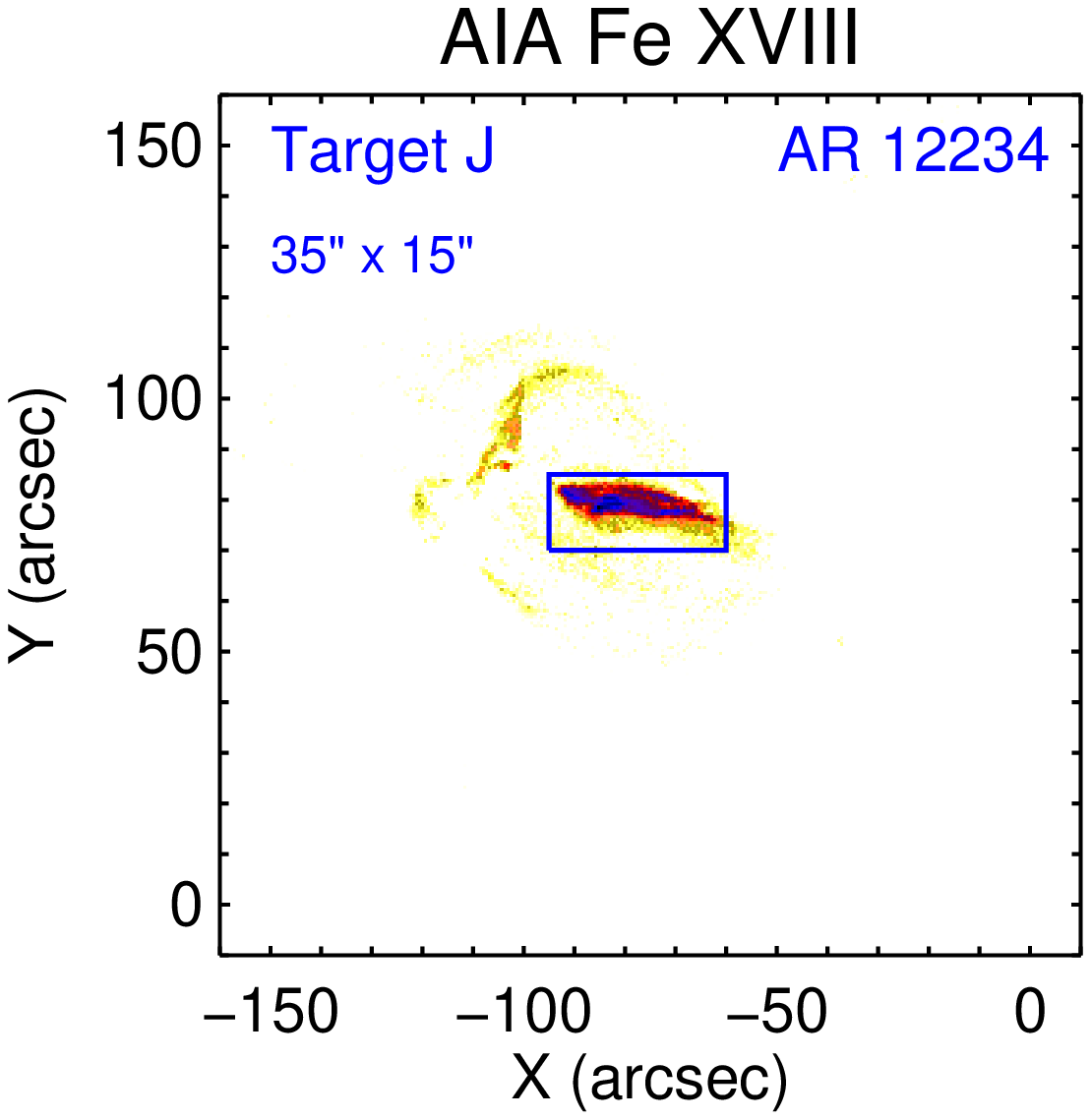}
\hspace{-1.5em}
\includegraphics[width=0.26\linewidth,trim ={3cm 0cm 3cm 1.65cm},clip]{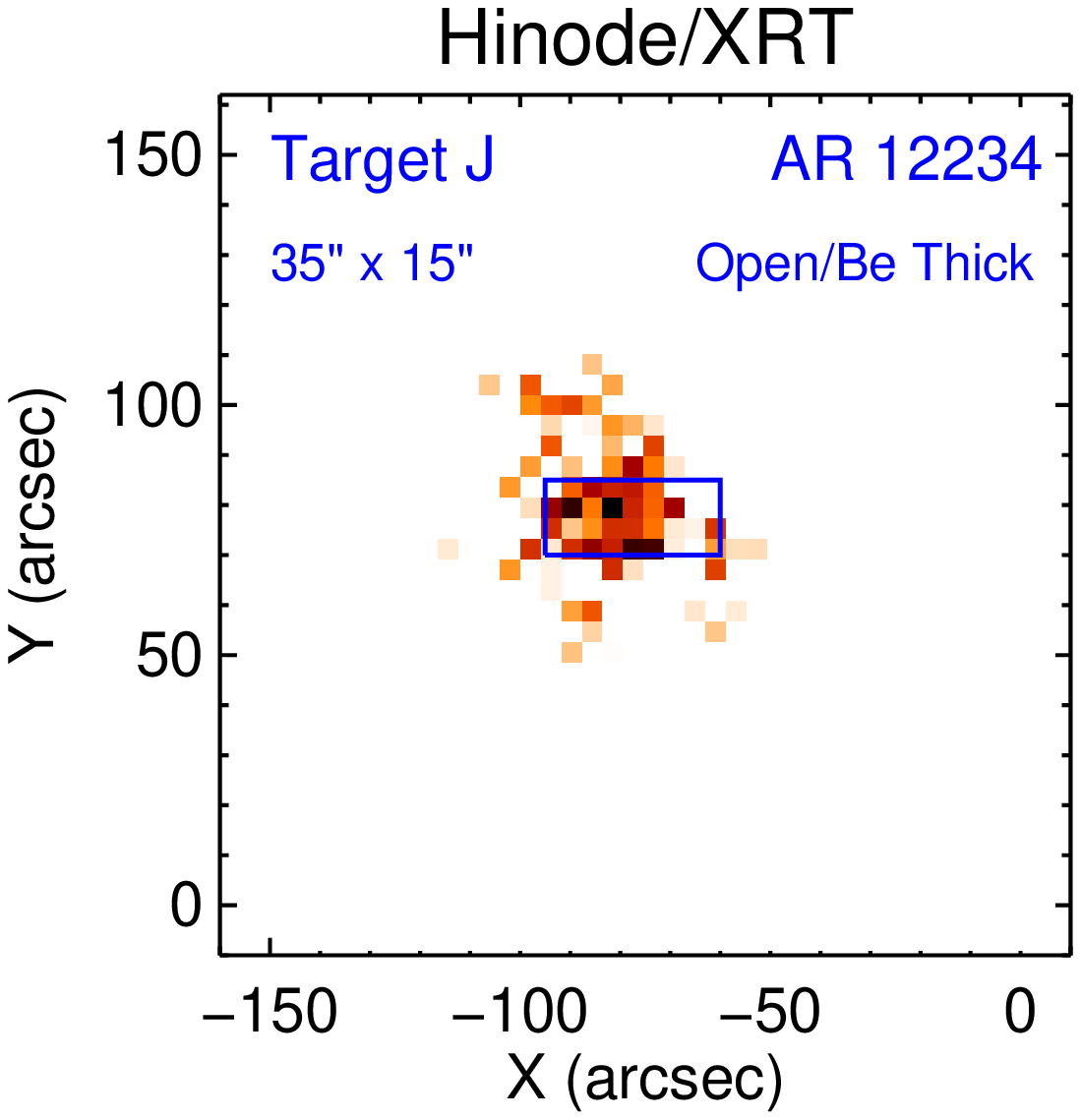}

\caption{Plots of AR12230 (Top) and AR12234(Bottom) during Target C and Target J of {\foxsi} observation. Images from {\aia} EUV channels are pre-flare emission subtracted.  For both ARs, the  loops are well observed in the hotter {\aia} 94 {{\AA}}, 131 {{\AA}}, FeXVIII and {\xrt} SXR channels. The bright loop regions used for the DEM analysis are overplotted as blue rectangles. It is to be noted the instruments have different pixel resolutions.}

\label{fig:aiaimage}
\end{figure}

We recovered the Fe XVIII contribution to the 94 {{{\AA}}} channel, which is a diagnostic for high temperature emission (${\approx}$ 7MK), from a  linear combination of {\aia} channels using the approach of \citet{Zanna2013}, as given by equation 1.

\begin{equation}
F(Fe XVIII) {\approx}F(94 \mbox{\AA}) - \frac{F(211 \mbox{\AA})}{120} - \frac{F(171 \mbox{\AA})}{450}
\end{equation}

where F(Fe XVIII) is the flux in the Fe XVIII line [DN s$^{-1}$ px$^{-1}$] and F (94 {{\AA}}), F (171 {{\AA}}), F (211 {{\AA}}), are the corresponding measured fluxes in the {\aia} 94 {{\AA}}, 171 {{\AA}}, and 211 {{\AA}} channels.  EUV and SXR images of AR12230 and AR12234 during Target C and Target J from {\aia} and {\xrt} are shown in
Figure \ref{fig:aiaimage} (Top and Bottom panels). The main loops of the ARs show brightening, indicating a large amount of hot plasma. The loop brightening is evident in the {\xrt} SXR channels and hotter EUV channels. We consider the flare area from the bright loop regions using Fe XVIII maps, which are 46 arcseconds ${\times}$ 28 arcseconds for AR12230 and 35 arcseconds ${\times}$ 15 arcseconds respectively, indicated by blue rectangular boxes in Figure \ref{fig:aiaimage}. Note that our flare area selection provides a conservative upper limit, as the bright FeXVIII pixels fill $>$40\% of the box for Microflare 1 and $>$ 65\% for Microflare 2.

\section{Temperature response functions}
\label{label:tempresp}
A  temperature response function represents the sensitivity of an instrument to detect plasma at different temperatures. It provides the expected count rate per pixel to be observed in a detector at a particular wavelength or energy band from a hot plasma at different isothermal temperatures. The temperature response function is determined by the photon transmission efficiency of the optics including windows/filters before reaching the detector and the detector's response to the incident photons and quantum efficiency. These factors also vary strongly as a function of wavelength/energy.

Here, we describe the temperature response functions of all the instruments, with a detailed explanation of the construction of the temperature response function for the {\foxsi} instrument.

\subsection{Temperature response for  {\aia} and {\xrt}}

The temperature response functions for the {\aia} EUV channels were obtained using \verb!aia_get_response.pro! with flags \verb!timedepend_date!, \verb!eve_norm! and \verb!chiantifix!. The response functions were constructed with version 6, using CHIANTI database v 7.1.3. Standard procedures were followed to construct the temperature response functions for {\xrt} using CHIANTI v 7.1.3 and   \verb!xrt_flux713.pro! (eg., \citet{Kobelski2014}). For this, we have considered coronal abundances taken from \citet{Feldman1992} and adopted the latest filter calibrations which incorporate the time-dependent contamination layer on the detectors \citep{Narukage2014}. While combining multiple instruments for DEM analysis, it is shown in \cite{Schmelz2015,Schmelz2016} and \citet{Wright2017} that the standard temperature response of XRT needs a cross-calibration factor, which is different from unity.  We also found a systematic excess in the prediction of XRT values in the first attempts at reconstructing a DEM. We therefore multiplied the {\xrt} responses by a factor of two to account for this cross-calibration, which is of the same order as quoted in previous literature.

\subsection{{\foxsi}'s Temperature response}

\begin{figure}[h]
\centering
\hspace{-3.0em}
\includegraphics[width=0.63\linewidth]{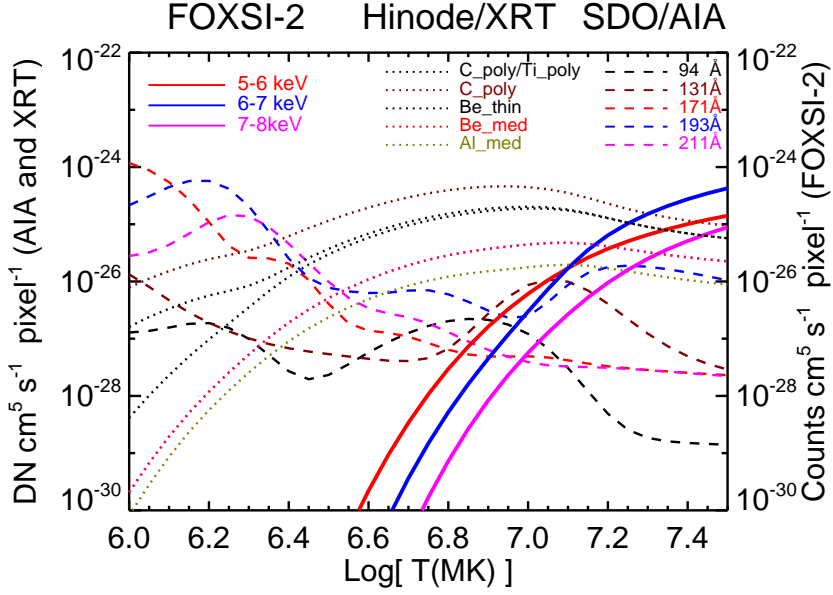}
\caption{The temperature response functions for {\foxsi} (D6 with 10 shell optics module (solid line), {\aia} EUV filters (dashed lines) and {\xrt} filters (dotted lines). The {\xrt} filter responses were multiplied by (${\times}$  2) for cross-calibration, which better matched the observed fluxes. It is to be noted that all the instruments have different pixel sizes and (in the case of AIA and XRT) DN  conversions, so absolute values between the instruments should not be directly compared. Rather, the plot conveys that the temperature sensitivity of {\foxsi} starts to increase above 5MK, while the other two instruments show a declining trend at those temperatures. This also shows a good overlap in the temperature sensitivity of all  the instruments, highlighting their complementarity.}
\label{fig:tresp}
\end{figure}
The temperature response curve for {\foxsi} was constructed from many discrete isothermal emission models ranging from 1MK up to 30MK in steps of  ${\delta}$ logT = 0.05. For each temperature, a synthetic X-ray emission photon spectrum was created using the CHIANTI database (v 7.1.3) \citep{Dere1997,Landi2013} with coronal elemental abundances. Each synthetic photon spectrum was then folded through the {\foxsi} instrument response (described below), to obtain a synthetic counts spectrum as a function of photon energy. The average counts integrated over 1 keV energy bins were obtained in the energy range 4 to 10 keV. Thus, we have a matrix of plasma temperatures and {\foxsi} energy bands for which we have the predicted count rates.  The temperature response function for each detector/optic module pair of {\foxsi} is  unique due to the differences in the effective area of the optic module and detection efficiency.

The instrument response for {\foxsi} was obtained using the ground calibration data for the optic modules and detectors. A thorough calibration of X-ray optic modules including the measurement of effective area was carried out using the stray light X-ray facility at the Marshall Space Flight Center \citep{Christe2016}. {\foxsi} detectors were calibrated using radioactive sources and we have measured the spectral resolution at different X-ray energies \citep{Ishikawa2011,Ishikawa2014,Athiray2017}.  The detection efficiency was calculated based on the transmission of X-rays through the optical path that includes thermal blanket material and filters, and absorption in the active volume of detector. Finally, the instrument response for a detector/optic pair was obtained by multiplying the effective area of an optic module with its corresponding detector efficiency. In our approach, we adopted a quasi-diagonal spectral response matrix, which also includes the detector's spectral resolution of ${\sim}$ 0.5 keV (FWHM) in the energy range 4 - 15 keV.

The temperature response functions for {\foxsi} (D6), {\aia}, and {\xrt} filters used in our  analysis are shown in Figure \ref{fig:tresp}. We note that all the instruments have different pixel sizes and DN conversions (in the case of AIA and XRT). Therefore, the absolute values between the instruments should not be directly compared. Three {\foxsi} energy bins were considered with the energy ranges 5 - 6 keV, 6 - 7 keV and 7 - 8 keV. The selection of energy bands for {\foxsi} was based on the criteria explained in section \ref{label:foxsisummary}. We also note that {\foxsi}'s temperature response  starts to increase above 5MK, whereas the other two instruments start dropping in their   sensitivity at those temperatures. Furthermore, {\foxsi} also shows a good overlap in the temperature sensitivity with the other two instruments. This allows to better constrain the slope of the high temperature emission.

\section{Combined DEM analysis of microflare emission}
\label{label:DEManalysis}
For the microflare DEM analysis, data from Target A, B, and C are considered for Microflare 1, while  data from Target J are considered for Microflare 2.   Therefore, for each Target we have ten channels (5 AIA; 3 {\fox}; 2 XRT) available for Microflare 1 and nine channels (5 AIA; 3 {\fox}; 1 XRT) available for Microflare 2. With the assumption that the observed coronal plasma is optically thin, the observed intensities (Y$_i$) in a wavelength/energy band are given by,

\begin{equation}
Y_i = R_{i,j} {\times}{\xi}(T_j)
\end{equation}
 where R$_{i,j}$ is the temperature response function for the {\it i} th filter channel and {\it j} th temperature bin, and  ${\xi}$(T)  is the line-of-sight DEM.

We adopted the Hinode-XRT method for the DEM recovery, which is a forward-fitting approach that uses a functional form of a spline implemented as {\it \verb! xrt_dem_iterative2.pro!} \citep{Golub2004}.  The best fit DEM distribution is identified iteratively using a non-linear least-squares method by comparing  the predicted and observed fluxes as a part of the minimization procedure. The significance of the best-fit solution is determined through Monte-Carlo (MC) runs, which are performed by varying the observed intensities by random values within the observed errors.  This method has been used in DEM fitting with SDO/AIA, Hinode/XRT and Hinode/EIS data \citep[e.g.][]{Golub2007, Winebarger2011, Wright2017, Ishikawa2017}.

The uncertainties in AIA data were calculated using \verb!aia_bp_estimate.pro! \citep{Boerner2012}. For {\xrt}, the photon statistical uncertainties were calculated based on \cite{Narukage2011}; non-statistical uncertainties were obtained using \verb!xrt_prep.pro!\citep{Kobelski2014,Narukage2011}. The statistical uncertainties for the {\foxsi} data were obtained for the respective duration after binning the photons by energy using the processed level 2 data.  We assumed a 10\% systematic uncertainty to the observed values and added them in quadrature with the statistical uncertainties.

\begin{figure}[h]

 \subfigure{\label{fig:a41}
 \hspace{-2.0em}

     \includegraphics[width=0.35\linewidth]{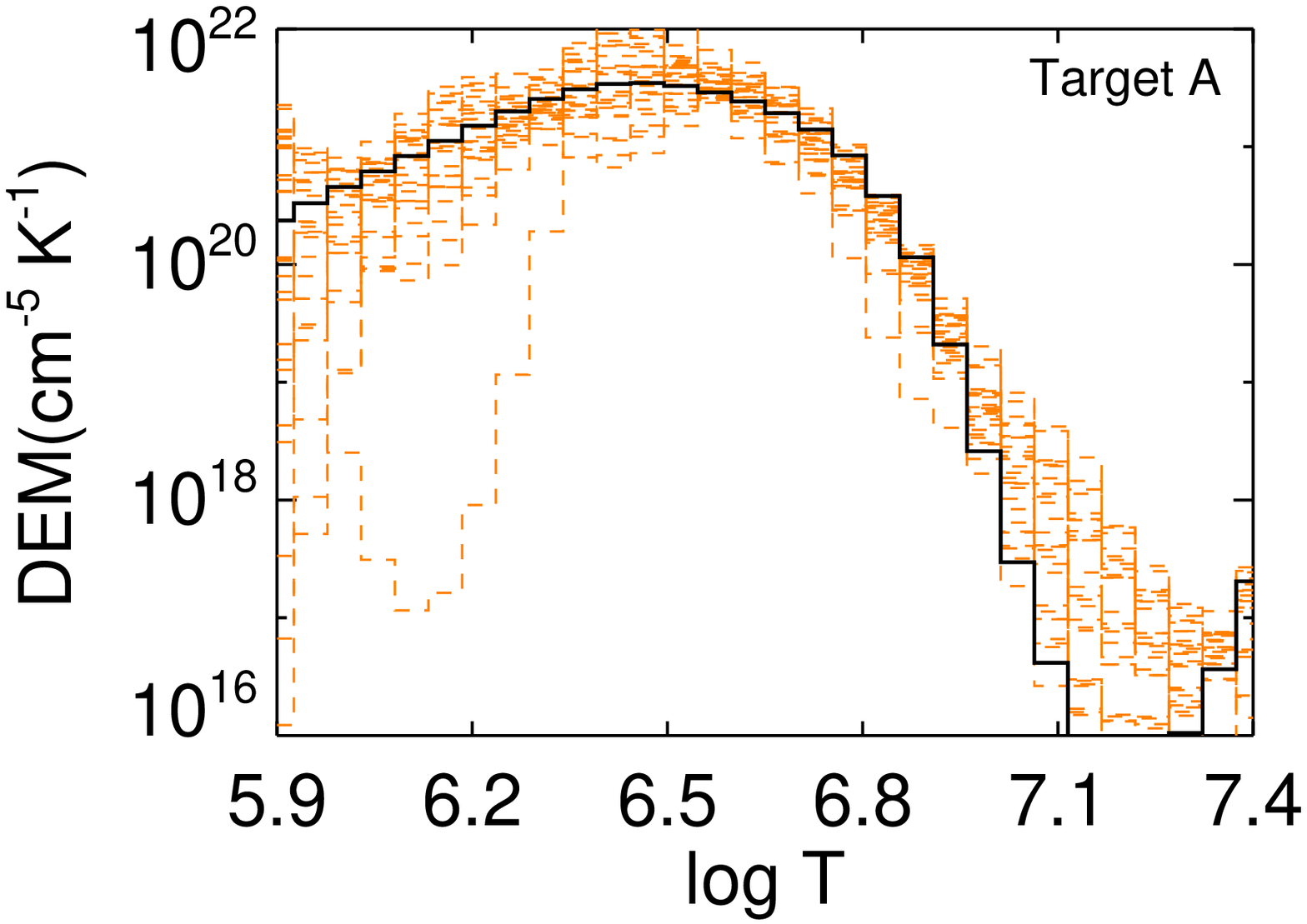}}
      \hspace{-1em}
      \vspace{-1em}
\subfigure{\label{fig:b41}

        \includegraphics[width=0.35\linewidth]{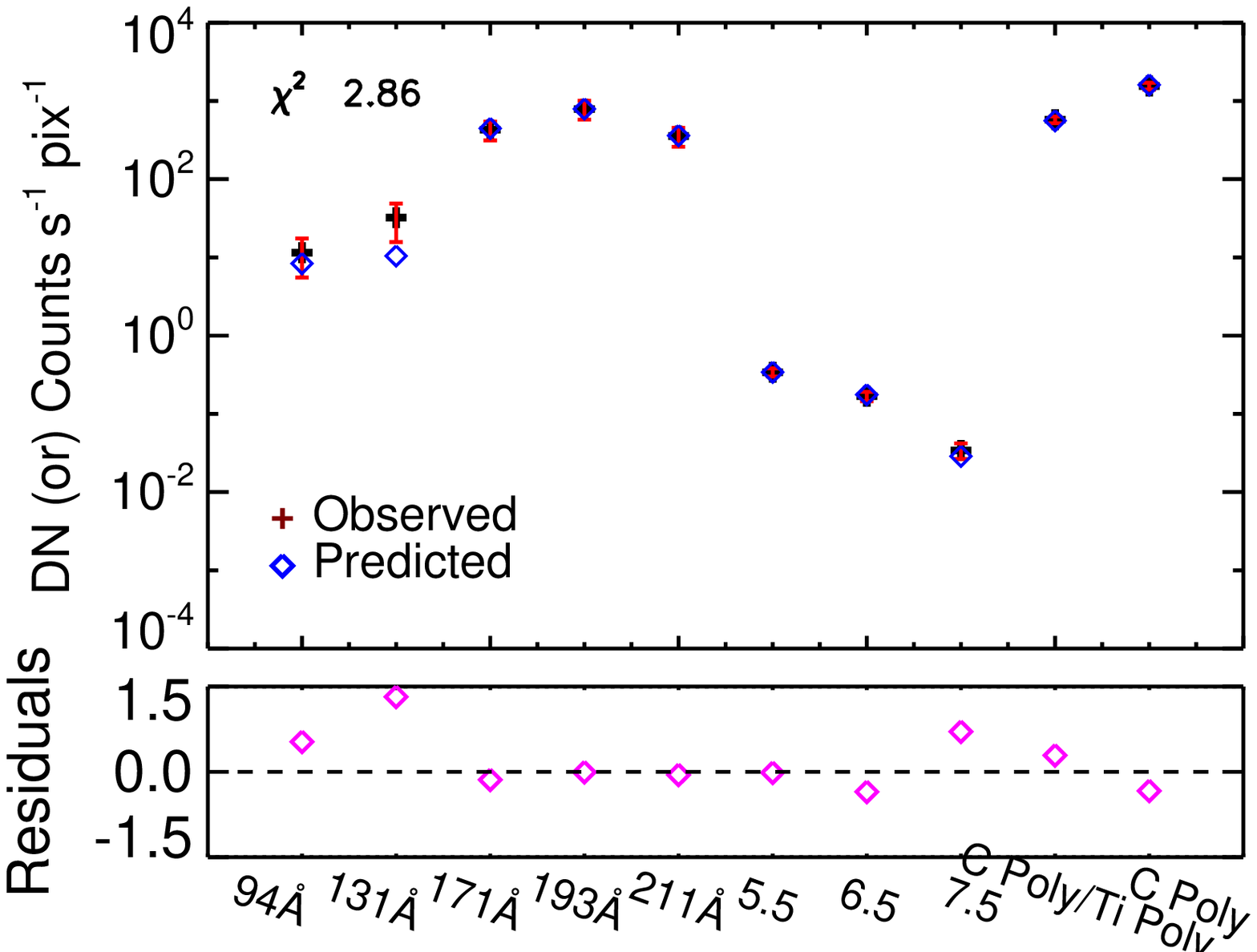}}
          \hspace{-1em}
\subfigure{\label{fig:c41}

          \includegraphics[width=0.35\linewidth]{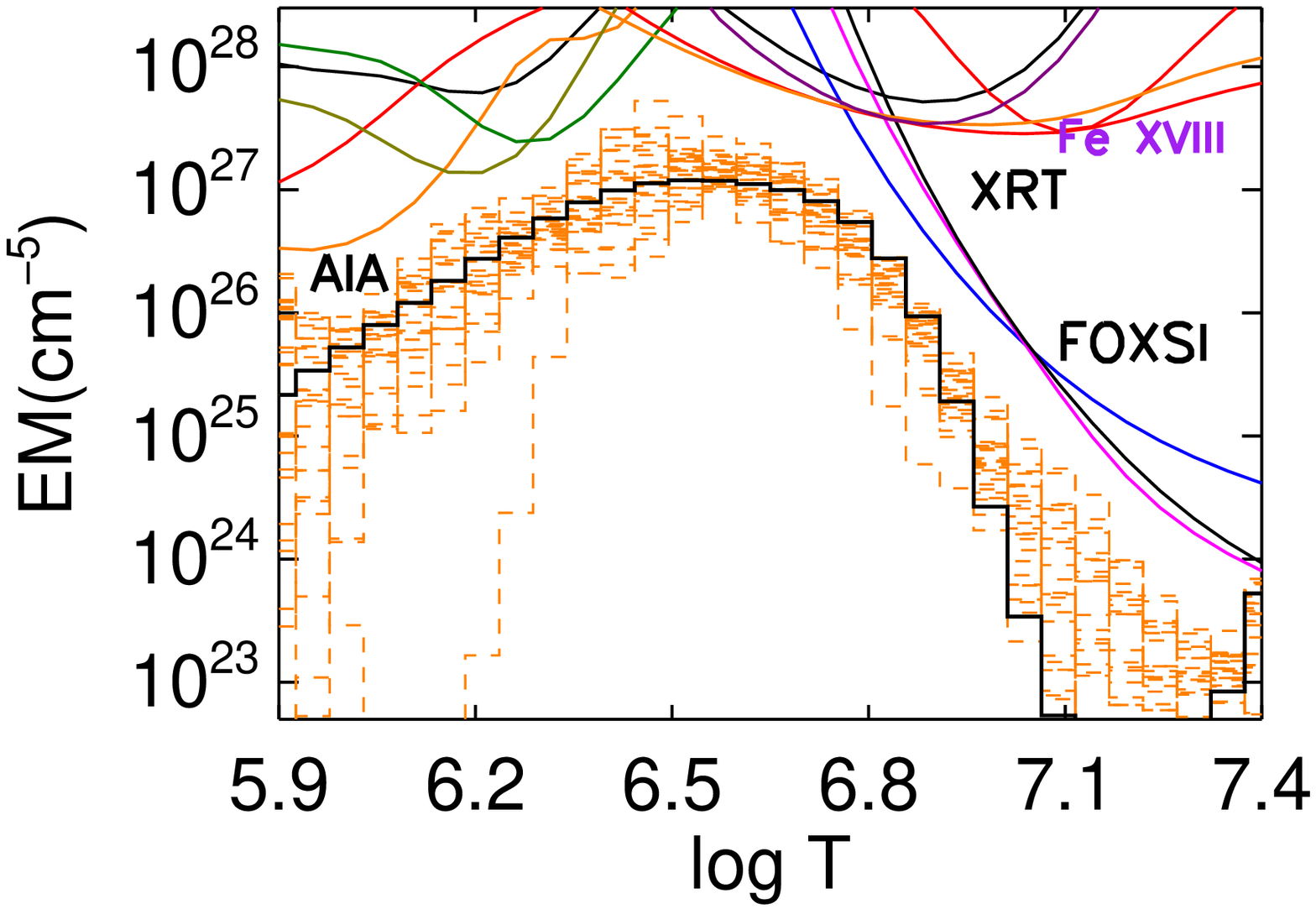}}
\subfigure{\label{fig:a42}
\hspace{-2.0em}

          \includegraphics[width=0.35\linewidth]{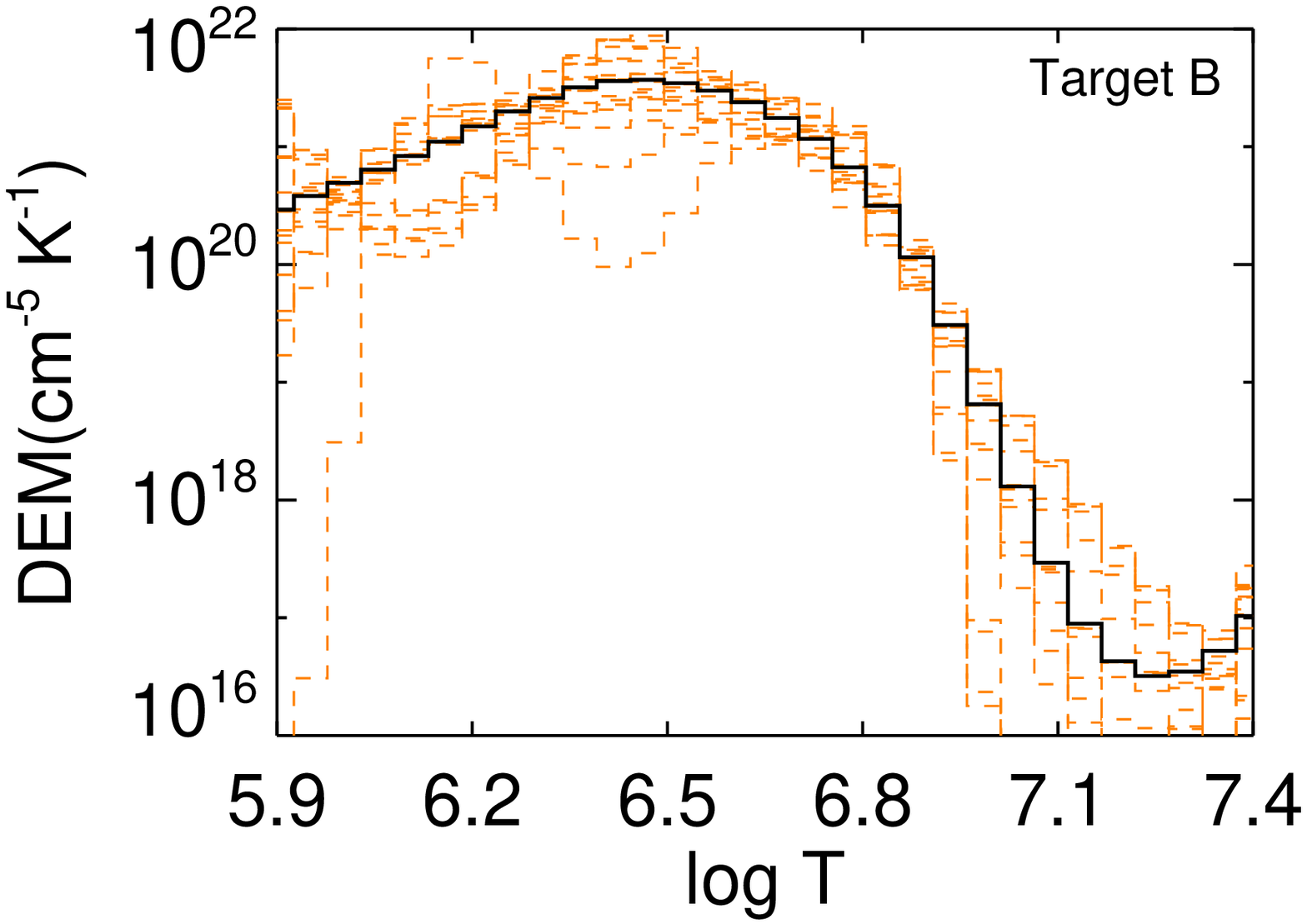}}
\subfigure{\label{fig:b42}
            \hspace{-1em}
        \includegraphics[width=0.35\linewidth]{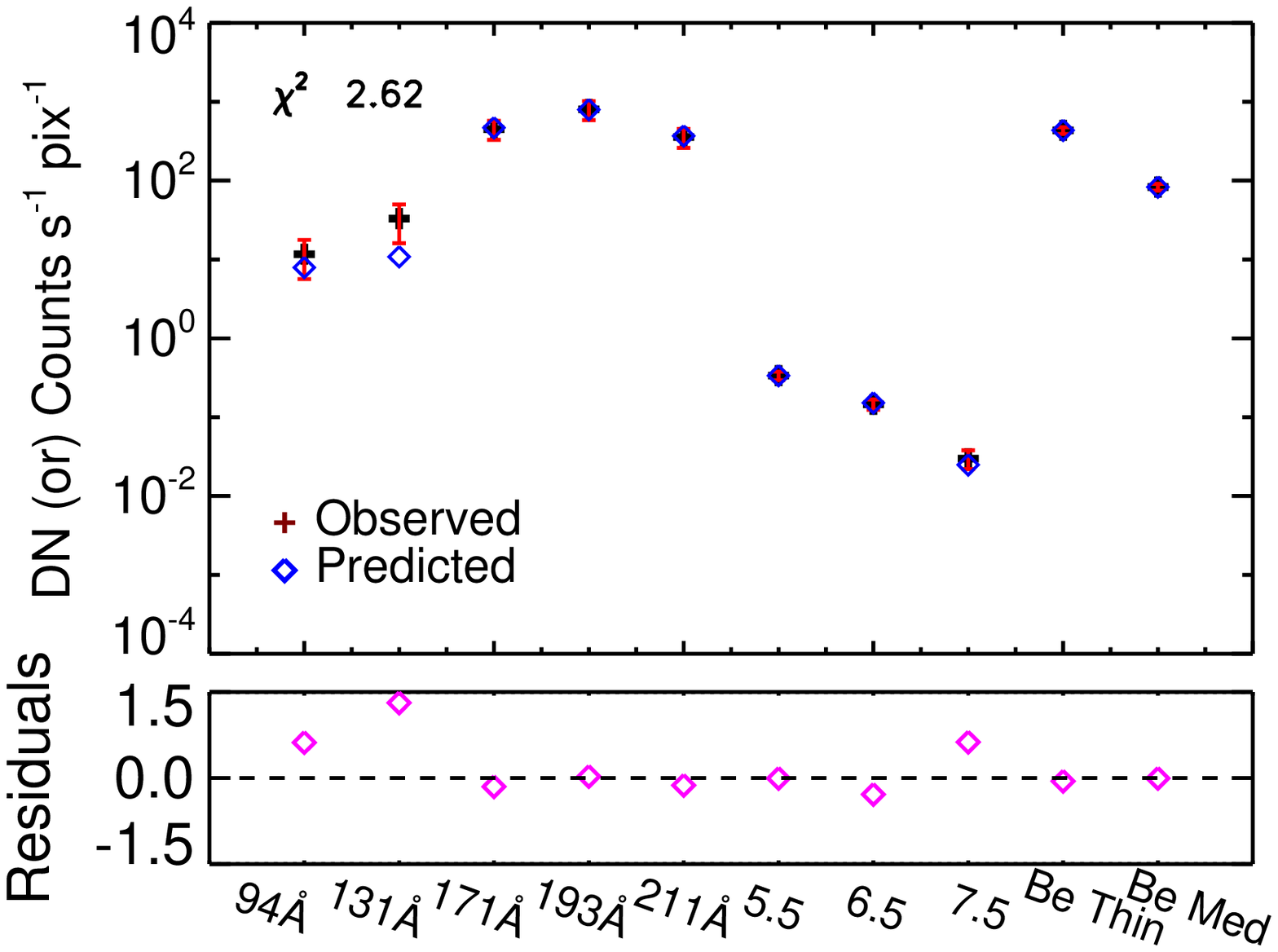}}
\subfigure{\label{fig:c42}
          \hspace{-1em}

    \includegraphics[width=0.35\linewidth]{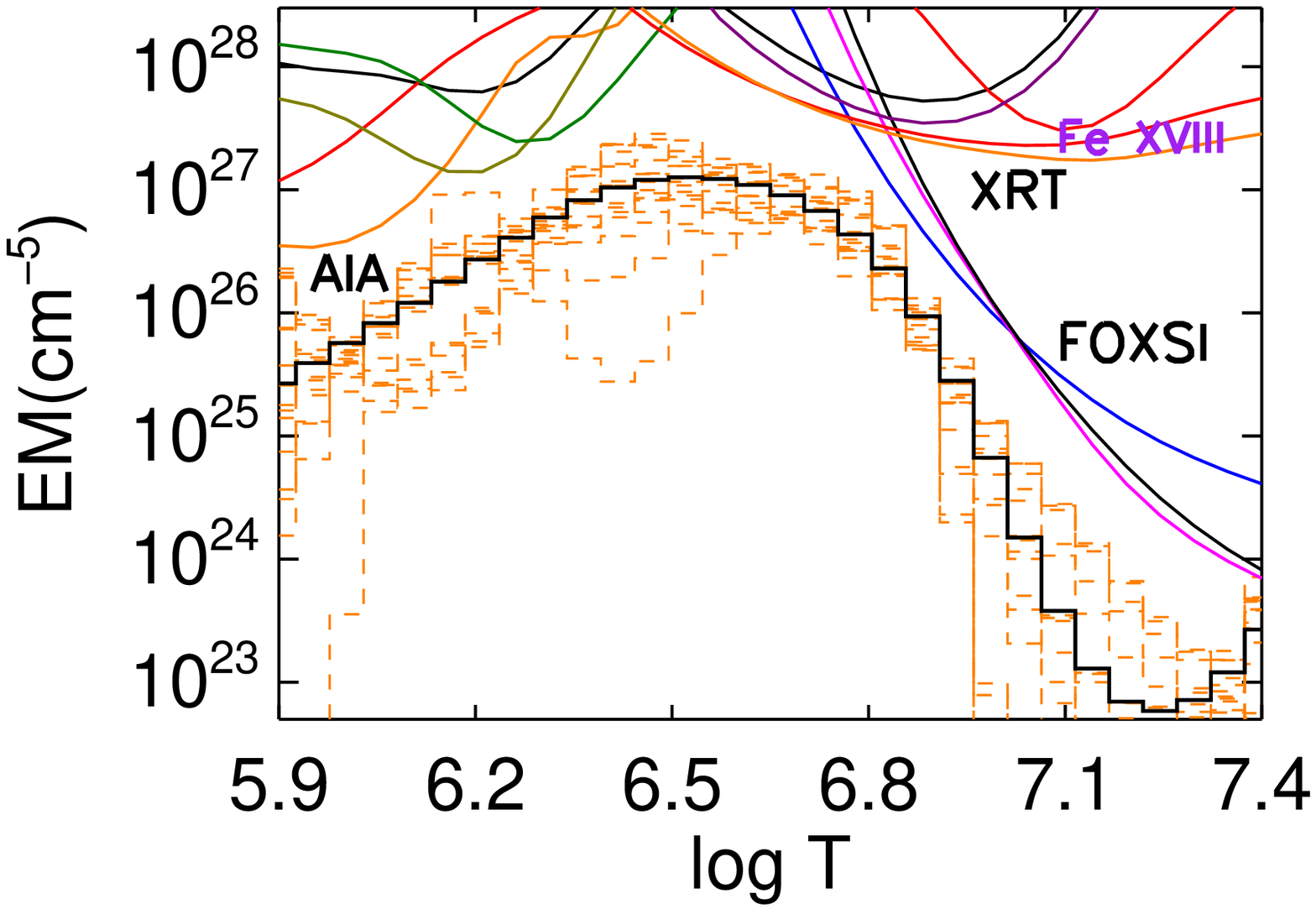}}
\subfigure{\label{fig:a43}
      \hspace{-2.0em}

         \includegraphics[width=0.35\linewidth]{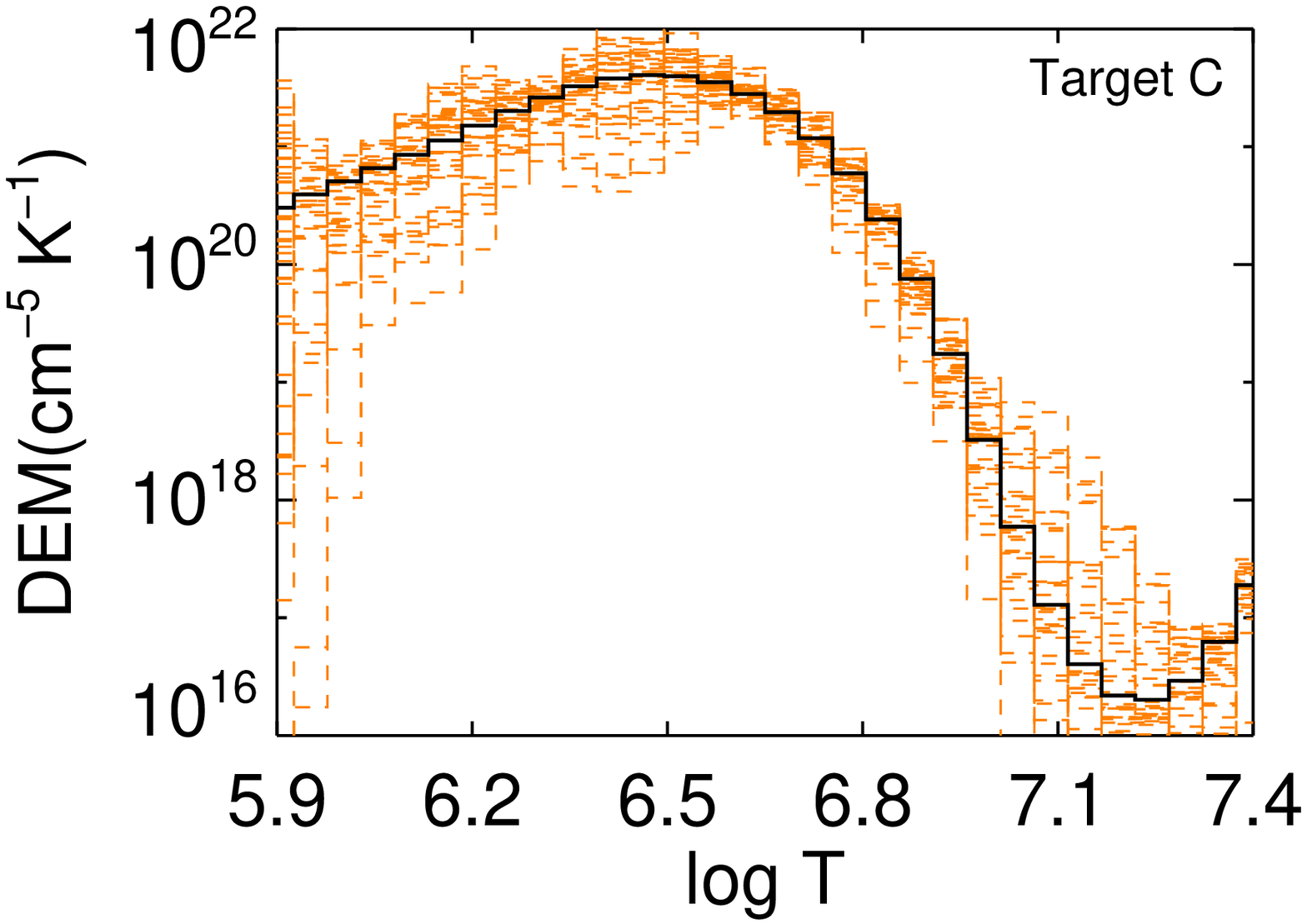}}
\subfigure{\label{fig:b43}
      \hspace{-1em}

    \includegraphics[width=0.35\linewidth]{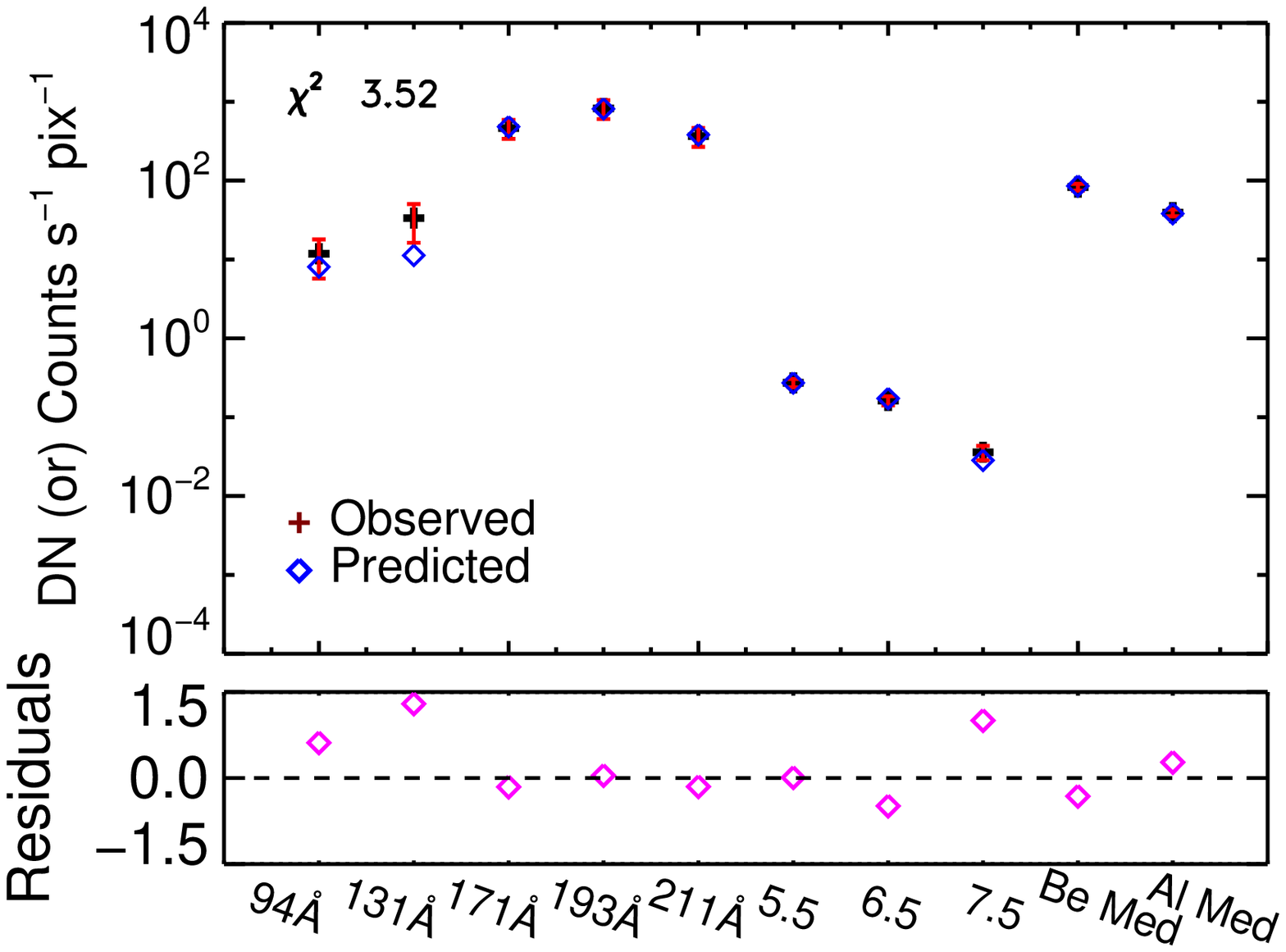}}
    \subfigure{\label{fig:c43}
          \hspace{-1em}

    \includegraphics[width=0.35\linewidth]{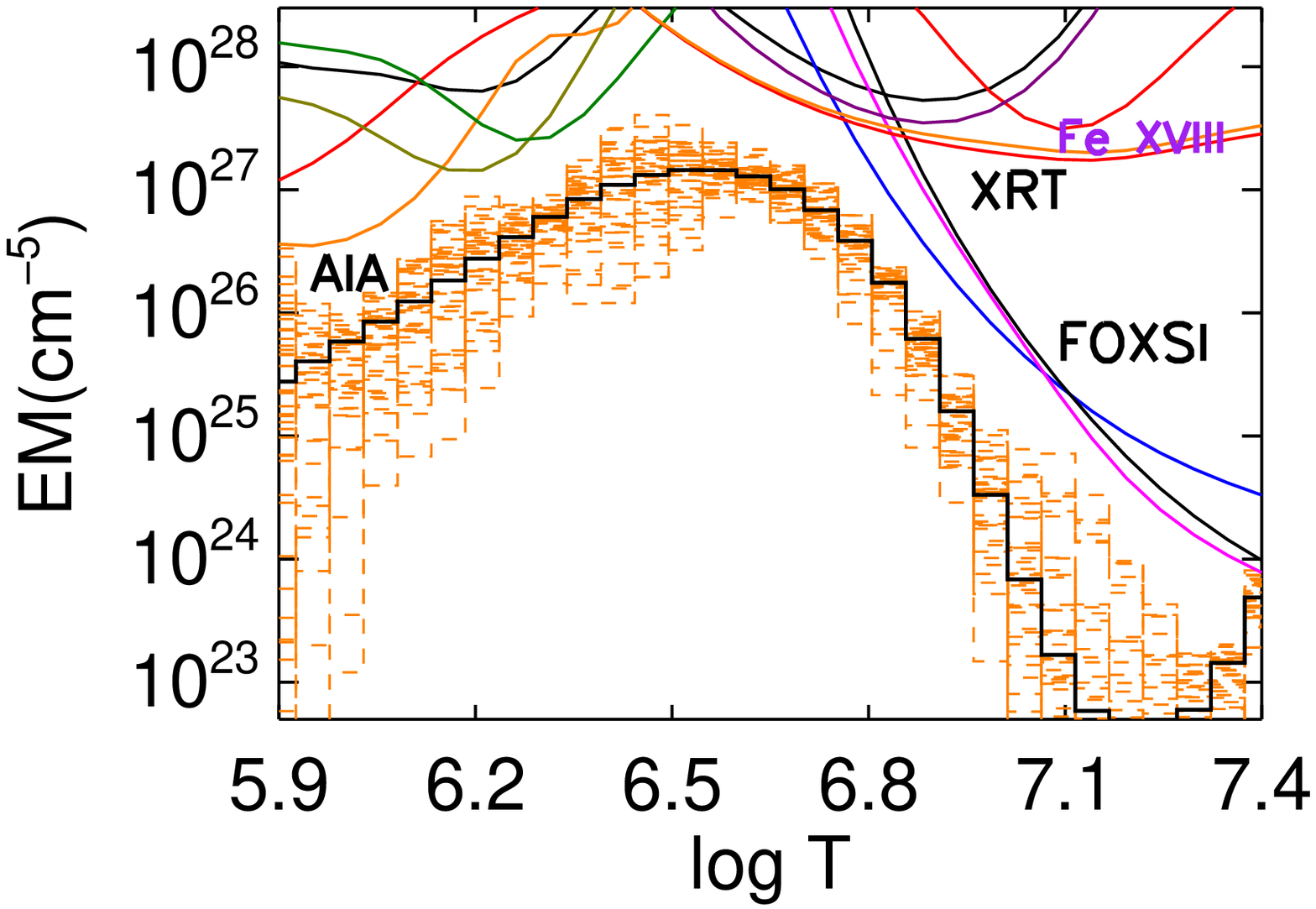}}
                \hspace{-1em}
\caption{(Left column) The DEM solutions obtained for Microflare 1 during Target A (Top row), Target B (Middle row) and Target C (Bottom row) using {\foxsi}, {\xrt} and {\aia} data. The best-fit solution for the observed fluxes is shown as a solid black line; selected MC solutions are shown as orange dashed lines. (Middle column) Comparison of observed and  best-solution predicted fluxes. A close agreement between the observed and predicted fluxes is apparent in the residuals in the bottom panels. The chi-square (${\chi}^2$) values correspond to the best-fit DEM solution. (Right column) The emission measure distributions (EMD) overplotted  with EM loci curves.}
\label{fig:DEMsolution}
\end{figure}

\begin{figure}[h]
 \hspace{-2.0em}
      \includegraphics[width=0.35\linewidth]{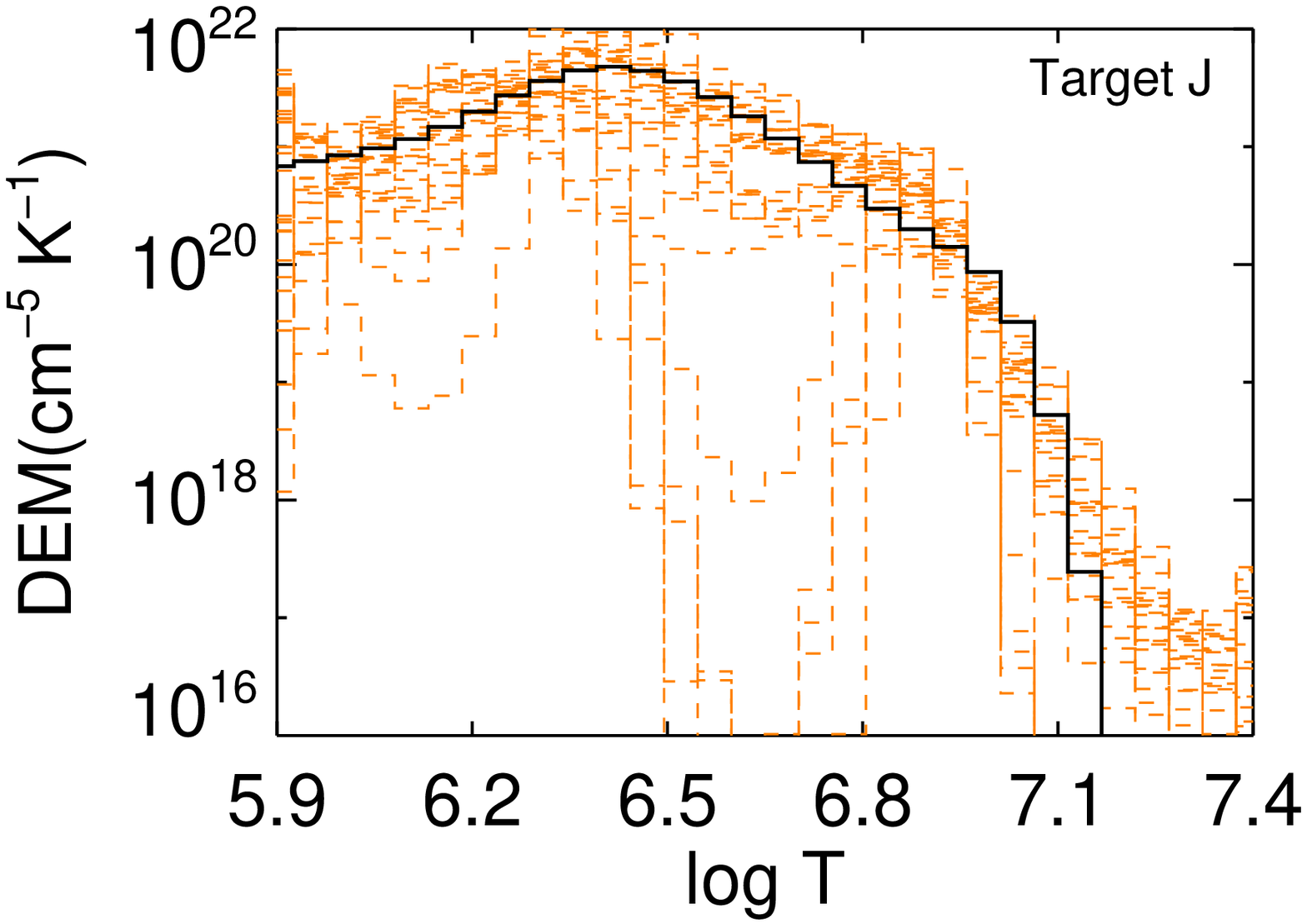}
      \hspace{-1em}
    \includegraphics[width=0.35\linewidth]{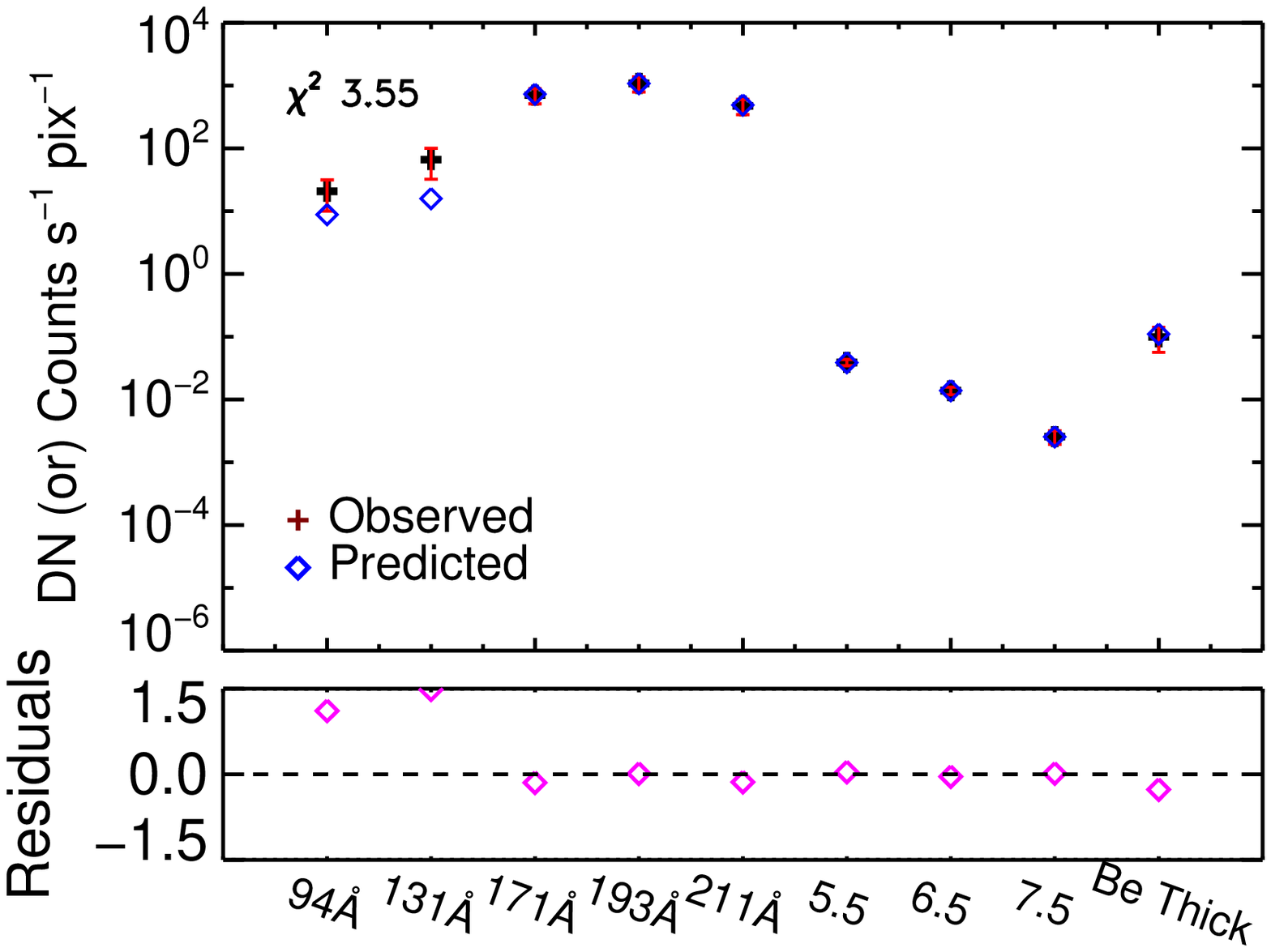}
          \hspace{-0.5em}
      \includegraphics[width=0.35\linewidth]{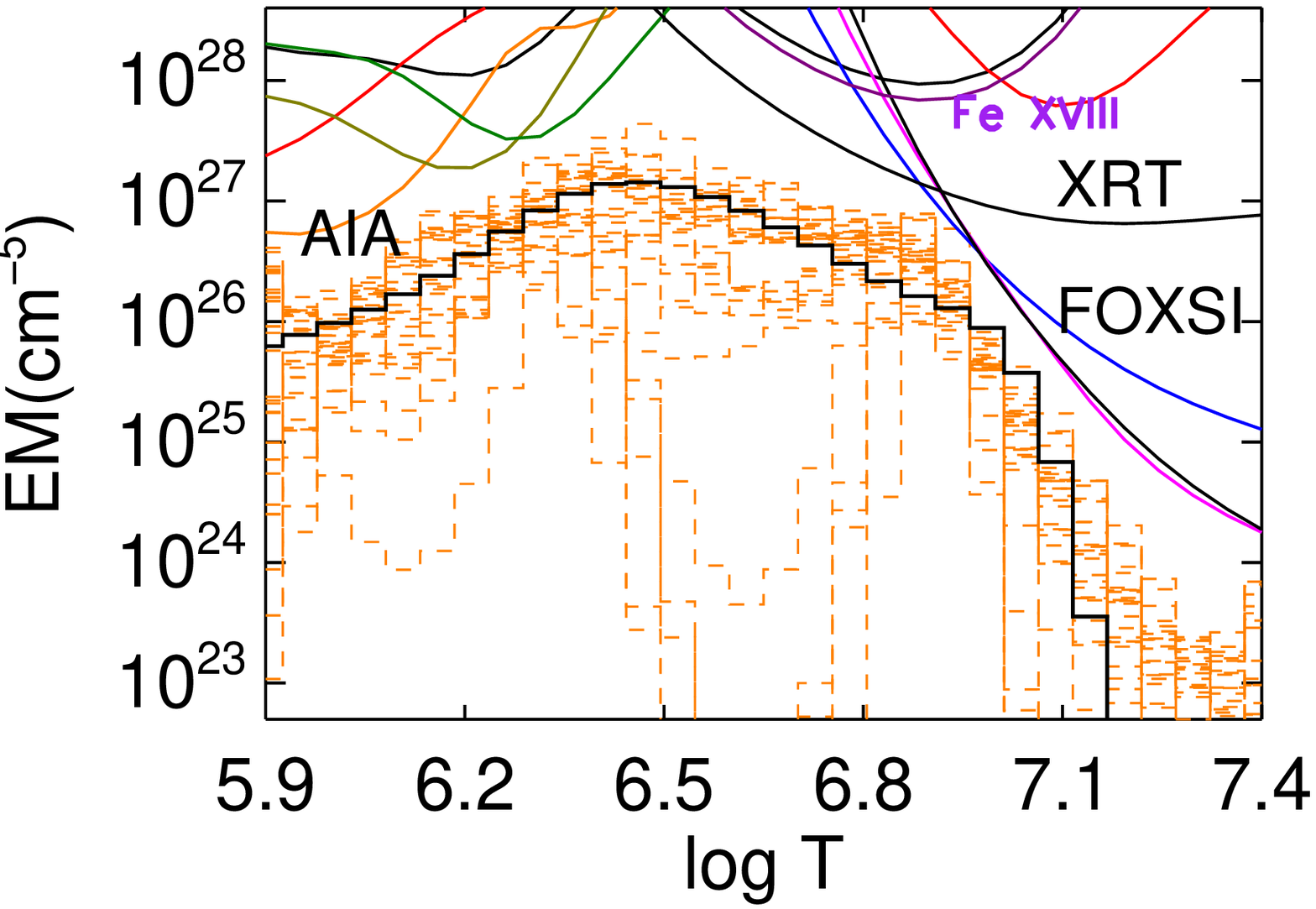}

\caption{(Left) The DEM solutions obtained for Microflare 2 during Target J using {\foxsi}, {\xrt} and {\aia} data. The best-fit solution for the observed fluxes is shown as a solid black line; selected MC solutions are shown as orange dashed lines. (Middle) Comparison of observed and best-solution predicted values. A close agreement between the observed and predicted fluxes is shown as residuals in the bottom panels. The chisquare (${\chi}^2$) values correspond to the best-fit DEM solution. (Right) The emission measure distribution (EMD) overplotted with EM loci curves of the instruments (labeled).}

\label{fig:flare2_DEMsolution}
\end{figure}

\begin{figure}[h]
\hspace{-1em}
\hspace{-0.5em}
\includegraphics[width=0.5\linewidth]{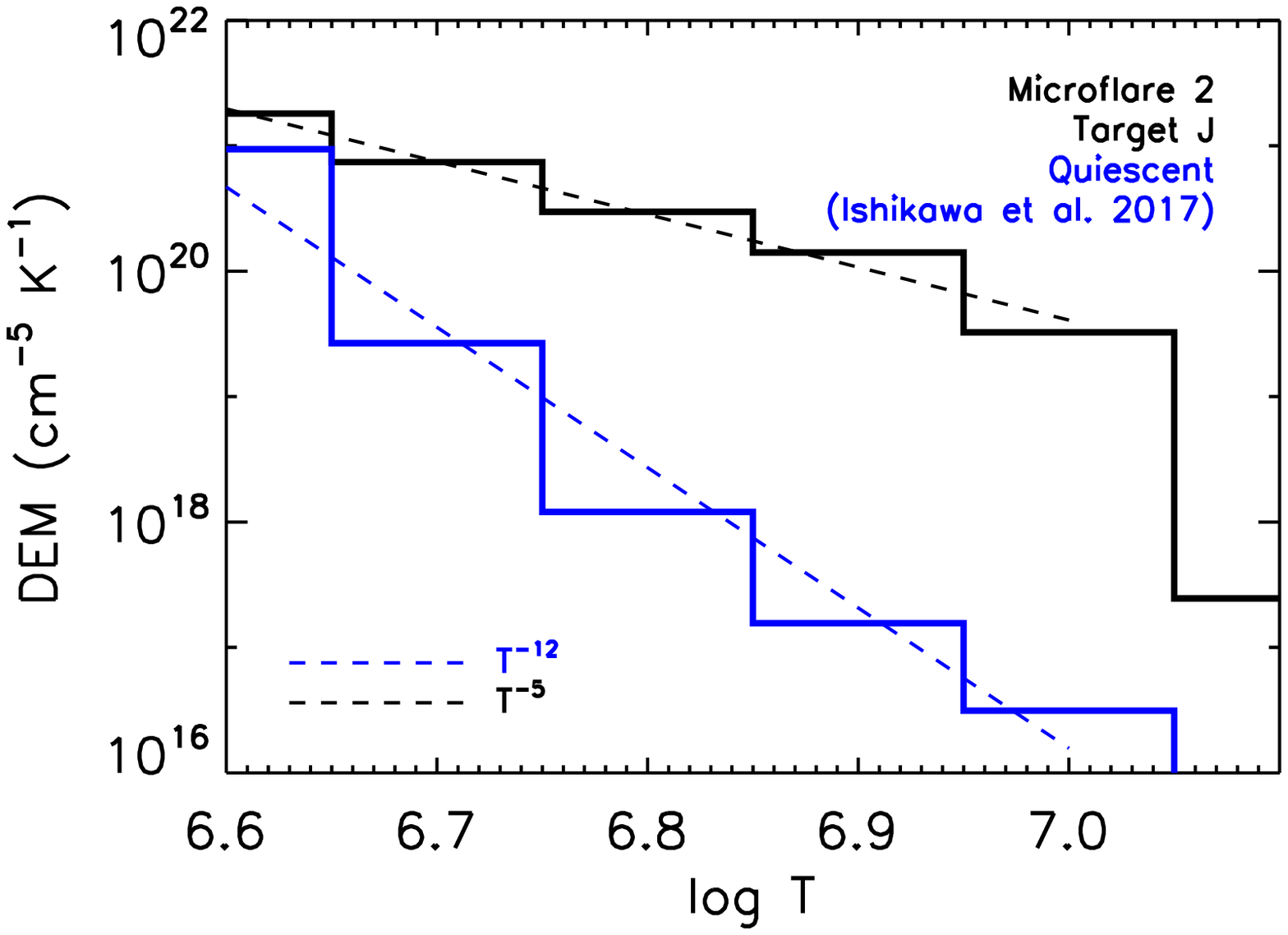}
\includegraphics[width=0.5\linewidth]{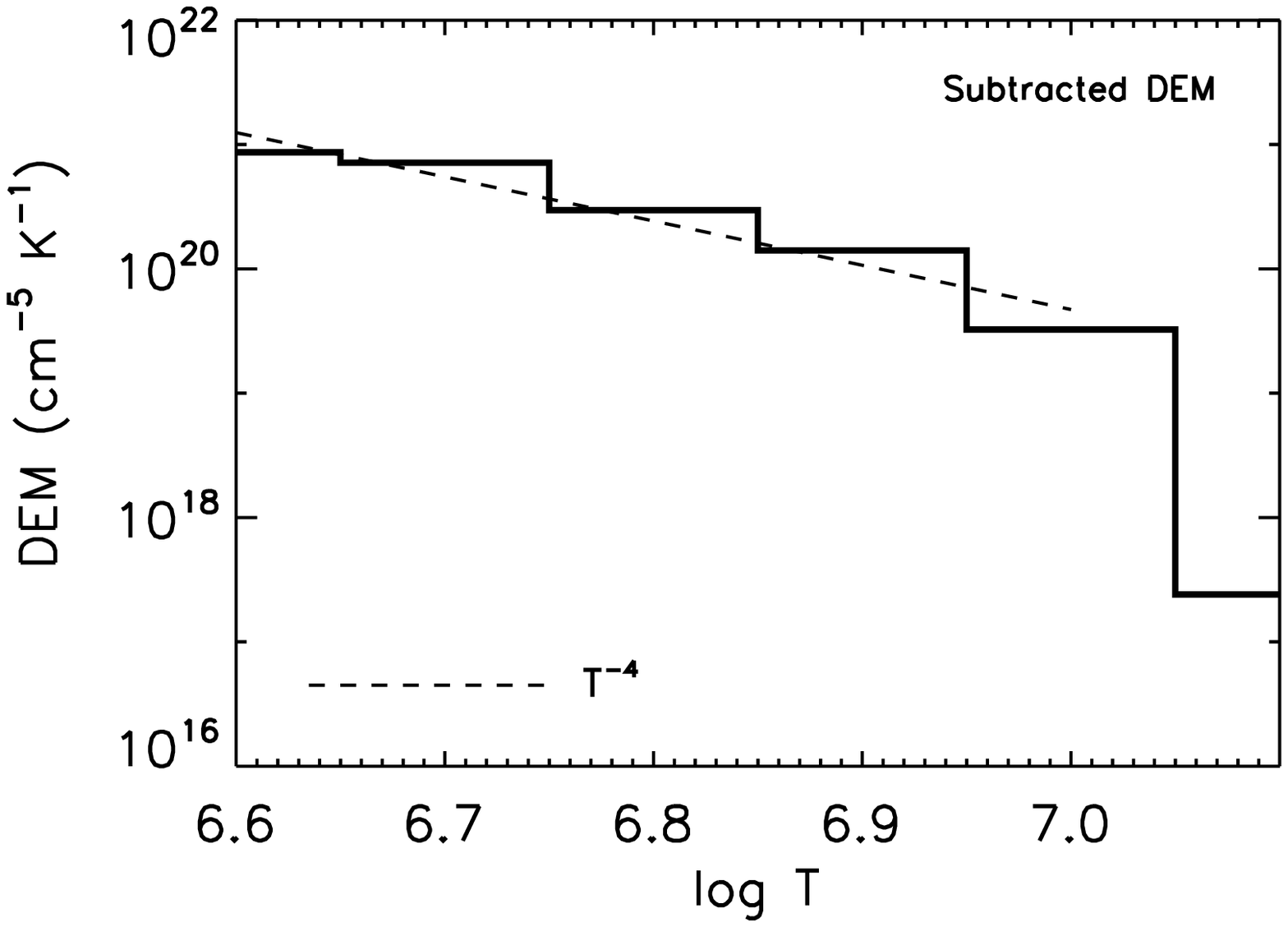}
\caption{Comparison of DEMs during flare (Target J) and non-flare times for AR12234. (Left). The quiet AR DEM for this region is taken from \cite{Ishikawa2017}. (Right) This plot shows the flare DEM subtracted from the non-flare AR DEM for Microflare 2. The  decrease in slope from quiet (${\propto}$ $T^{-12}$) to flare (${\propto}$ $T^{-4}$) phase shows the presence of excess hot plasma > 4MK contributed from the microflare.}
\label{fig:EMcomparison_internal}
 \end{figure}

The results of  the DEM analysis for Microflare 1 and Microflare 2 are shown in Figures \ref{fig:DEMsolution} and \ref{fig:flare2_DEMsolution}. The rows in Figure \ref{fig:DEMsolution} corresponds to Target A, B, and C. The resulting DEM solutions are shown in the left panels of Figures \ref{fig:DEMsolution} and \ref{fig:flare2_DEMsolution}. The best-fit DEM solution is denoted by a solid black curve. We performed 100 MC runs and considered solutions with chi-square values less than 2 times the best-fit solution chi-square value, which are shown as dashed orange lines. The spread in the MC solutions is an indicator of how well the solutions are constrained at a particular temperature range. The resulting DEMs peak around 3MK and extend beyond 10MK. The slope of the high temperature emission above 5MK is well determined by including the {\foxsi} HXR data. The goodness of the best-fit DEM solution is computed by comparing the predicted and observed fluxes with uncertainties as shown in the middle panels. A close agreement between the observed and predicted fluxes is indicated  as  residuals in the bottom of the middle panels. The emission measure distributions (EMDs) are shown in the right panels of Figures \ref{fig:DEMsolution} and \ref{fig:flare2_DEMsolution}. The EMDs are obtained by multiplying the DEM distribution with $T {\delta}logT$. Overplotted are the EM loci curves of the instrument channels, which provide upper limits for the EM at different isothermal temperatures. We also show the Fe XVIII loci curve (purple), which is closer to the DEM solution above 4MK than AIA 94~{\AA}. The response function for Fe XVIII was obtained using the linear combination of temperature responses of AIA channels \citep{Zanna2013}. This indicates a dominant cold component to the 94~{\AA} channel.  The true EM solution satisfies the expectation  to lie below the loci curves. It is clearly evident that including HXR data from {\foxsi} can better constrain the high temperature emission than using the instruments {\aia} and {\xrt} alone. We also note that {\foxsi} loci curves closely intersect at temperatures around 10MK. A similar range of temperatures is obtained from isothermal spectral fits, described in Paper II. Note that although multi-wavelength data indicate a multi-thermal DEM, this would not be apparent from the HXR data alone.

To determine how much emission exclusively comes from the microflare, we compare the DEM distributions during flare and non-flare times. For the first time we have simultaneous observation of microflaring ARs available in HXRs, SXRs and EUV for the quiescent and flare time intervals. Although Microflare 1 shows less HXR emission during Target J, we do not have compelling evidence to unambiguously claim that for  quiescent emission. For Microflare 2, we considered the quiet AR DEM established in \cite{Ishikawa2017}.  Figure \ref{fig:EMcomparison_internal} (Left) compares the flaring and non-flaring DEMs of AR12234 (Microflare 2). The steep DEM power-law relation ${\propto}$ $T^{-12}$ determined by \citep{Ishikawa2017}, between log T = 6.6 to 7.0, reduces to ${\propto}$ T$^{-5}$ during the flare phase. Figure \ref{fig:EMcomparison_internal} (Right) shows the microflare DEM subtracted from the quiet DEM with ${\propto}$ $T^{-4}$. The decrease in slope from quiet to flare phase indicates a clear presence of excess hot plasma above 4MK is chiefly contributed from the microflare.  These results highlight the sensitivity of {\foxsi} by detecting high-temperature plasma that are orders of magnitude lower than the detection limits of {\xrt} and {\aia}. The equivalent Geostationary Operational Environmental Satellites ({\goes}-15)  flare classes determined using the best-fit DEM solution yield A0.14 for Microflare 1 and A0.3 for Microflare 2, which are consistent with isothermal estimates (Paper II) .

\section{Thermal energy estimates}
\label{sec:Ten_estimates}
Using the best DEM solution, we calculate the thermal energy released based on \cite{Inglis2014}, which is given as:

\begin{equation}
U_{th} = 3k_BV^{\frac{1}{2}} \frac{{\int}_T T{\xi}(T)dT}{\sqrt[]{{\int}_T {\xi}(T)dT}} (erg)
\end{equation}

where $k_B$ is the Boltzmann constant. We assumed a filling factor of 1 and volume of the emission V = A$^{1.5}$, where A is the flare area. Therefore, the thermal energy estimates are upper limits as we have used conservative flare area selection described in Section \ref{label:aiaxrtsummary}. ${\xi}$(T) = $n_e^2 dV/dT$ is the DEM solution in the units of $cm^{-3} K^{-1}$. Table \ref{tab:thermalenergy} summarizes the thermal energy estimates calculated from DEM, ignoring any energy losses during heating. Thermal energy estimates computed using isothermal approximation are given in Table \ref{tab:thermalenergy} (Column 6) for comparison. This clearly shows the systematic underestimate of the thermal energy released when the isothermal approximation is used. This is consistent with the results obtained for the study of larger flares  (M and X-class) observed with {\aia} \citep{Aschwanden2015}. Our analyses show that the multi-thermal DEM provides more comprehensive thermal energy estimates than the isothermal approximation. For these microflares, multithermal DEM yields up to ${\sim}$ 4 times higher thermal  energy than the isothermal estimates.

\setcounter{table}{2}
\begin{table}[h!]
\renewcommand{\thetable}{\arabic{table}}
\centering
\caption{Thermal energy estimates of microflares observed during  {\foxsi} using the multi-thermal DEM analysis.} \label{tab:thermalenergy}
\begin{tabular}{cccccc}
\tablewidth{0pt}
\hline
\hline
Flare&Targets&Start & End & Multi-thermal DEM  & {Isothermal} \\
&&(UT) & (UT) & E$_{th} {\times}(10^{28}$ erg) &E$_{th} ({\times}10^{28}$ erg)   \\
\hline

1&A&19:12:42 & 19:13:14 & 5.1 $^{+0.7}_{-0.2}$  & 1.4 $^{+0.2}_{-0.2}$ \\
1&B&19:13:18 & 19:13:42 & 4.9 $^{+0.4}_{-0.4}$ & 1.5 $^{+0.2}_{-0.2}$ \\
1&C&19:13:47 & 19:14:25 & 5.1 $^{+0.6}_{-0.6}$ & 1.2 $^{+0.1}_{-0.1}$  \\
2&J&19:18:51 & 19:19:23 &1.6 $^{+0.6}_{-0.7}$ & 1.0 $^{+0.1}_{-0.1}$\\

\hline
\end{tabular}
\end{table}

\section{Discussion  and Summary}
\label{label:summary}

In this paper we presented the coordinated observations of two sub-A class microflares jointly observed by {\foxsi}, {\xrt} and {\aia}. These observations provide a unique opportunity to investigate small scale energy releases that were too faint to be captured in the {\rhessi}  and {\goes} flare catalogs. Significant HXR emission observed in {\foxsi} above 5 keV indicates the presence of  high temperature plasma up to 10MK. Concurrent brightening observed in the {\xrt} and {\aia} channels clearly indicate a multi-thermal plasma. We utilized the high sensitivity HXR data from {\foxsi} and performed a combined DEM analysis with {\xrt} and {\aia}, which together compose a good overlap in temperature sensitivity. The resulting microflare DEMs peak around 3MK and exhibit significant emission above 5MK. The coordinated {\foxsi} observations produce one of the few definitive measurements of the distribution and the amount of plasma above 5MK in microflares.

It has been established that existing solar instruments in the EUV and SXRs cannot precisely measure the low emission measure, high temperature plasma. Specifically, \citet{Winebarger2012} determined that there exists a “blind spot” in temperature-emission space for {\xrt} and EIS; these instruments cannot detect plasma with temperatures higher than 6MK and emission measures lower than 10$^{27}$ cm$^{-5}$. HXR measurements chiefly observe the bremsstrahlung continuum, whose emission is not heavily  affected by ionization timescales. However, both AIA and XRT channels are sensitive to line emissions, which might show up with a delay after the energy release due to the timescales involved in ionizing the plasma. Thus, continuum observations in HXRs can be well suited to determine the high temperature content resulting from the instantaneous impulsive heating \citep{Bradshaw2011}.

The DEMs of the same ARs were investigated by \cite{Schmelz2016} using {\xrt} and {\aia} alone, considering them to be a quiet, non-flaring ARs. We  note that the  flares do not  show up distinctly in the  XRT time profiles, while they  stand out  clearly in {\foxsi} time profiles. Their resulting DEM peaks around 3MK and has the same orders of magnitude of plasma content below 4MK, as in our analysis. However, the DEM results differ significantly from ours at high temperature end ($>$5MK). The recovered DEM by \cite{Schmelz2016} overestimates the amount of plasma at high temperatures, which would produce more HXR emission than observed. This shows the limitations of the instruments used (without {\foxsi}) to constrain the high temperature emission. We also note that they considered a small area in the core of the ARs and averaged data over a large time range (${\sim}$ 1 hour) in their analysis. In contrast, we used shorter {\foxsi} time intervals with a larger area to cover all the HXR emission. By including high sensitivity {\foxsi} HXR data we were able to determine a well-constrained DEM distribution above 5MK. Additionally, for the first time we have simultaneous observation of microflaring ARs available in HXRs, SXRs and EUV for the quiescent and flare time intervals. By comparing the DEMs during quiet and flaring phases for AR12234, we found that EM slope between logT = 6.6 to 7.0 decreases from ${\propto}$ T$^{-12}$ to ${\propto}$ T$^{-4}$ during flare times. This indicates a clear contribution of hot emission $>$ 5MK coming from the microflare.
We demonstrated that including HXR data from {\foxsi} can better constrain the high temperature emission than using the instruments {\aia} and {\xrt} alone.  The emission measure determined from {\foxsi} is lower than 10$^{26} cm^{-5}$ for temperatures higher than 5MK. Such faint emission at those temperatures can not be well-constrained by the other two instruments.

Using multi-thermal DEMs,  we  determined the comprehensive thermal energy estimates for the microflares. These results will provide significant observational evidence for coronal heating models.  A systematic DEM study of tiny micoflares with high sensitivity HXR measurements can help us to better understand the population of flare-frequency distribution and its contribution to coronal heating. In the future, we look forward to more high sensitivity coordinated observations from  missions such as the {\magixs} sounding rocket experiment \citep{doi:10.1117/12.856793,doi:10.1117/12.2313997,doi:10.1117/12.2232820,champey2019}, which has a good diagnostic capability for high-temperature, low-emission measure plasma \citep{Athiray2019}.

\section{Acknowledgements}
The {\fox} rocket team gratefully acknowledges NASA's LCAS program for funding
the experiment, specifically NASA grants NNX08AH42G and NNX11AB75G.  The team is
grateful to JAXA/ISAS for the donation of detectors, ASICs, and abundant
expertise, and to the NSROC teams at WSMR and Wallops for the excellent
operation of their systems. CHIANTI is a collaborative project involving NRL
(USA), RAL (UK), and the following Universities: College London (UK), of
Cambridge (UK), George Mason (USA), and of Florence (Italy).

\bibliographystyle{aasjournal}
\bibliography{sample}



\end{document}